\documentclass[11pt,article, onecolumn,draft]{IEEEtran}
\usepackage{times,amsfonts,amssymb,amsmath}
\usepackage{graphicx,url,bm,comment,subfigure}
\usepackage{epsfig}
\usepackage{color}

\def\BibTeX{{\rm B\kern-.05em{\sc i\kern-.025em b}\kern-.08em
 T\kern-.1667em\lower.7ex\hbox{E}\kern-.125emX}}

\newcommand{\profof}[1]
{

 \noindent
 {\bf Proof of #1.\\ }
 }

\newcommand{\EE}{\mathop{\mathbb{E}}\limits} 
\newcommand{\RR}{\mathbb{R}} 
\newcommand{\NN}{\mathbb{N}} 

\newcommand{\mems}{{\rm mem}}
\newcommand{\mem}{\nabla G{\rm mem}}
\newcommand{\imp}{\nabla G{\rm imp}}
\newcommand{\gmax}{\nabla G{\rm max}}
\newcommand{\Prob}
{
{\rm Pr}
}

\DeclareMathOperator*{\argmax}{arg\,max}

\newcommand{\Bc}{{\cal B}}
\newcommand{\Cc}{{\cal C}}
\newcommand{\Dc}{{\cal D}}
\newcommand{\Ec}{{\cal E}}
\newcommand{\Fc}{{\cal F}}

\newcommand{\Hc}{{\cal H}}

\newcommand{\Lc}{{\cal L}}

\newcommand{\Sc}{{\cal S}}

\newcommand{\Xc}{{\cal X}}

\newcommand{\DX}{{\cal D}_{\cal F}(X_t)}
\newcommand{\DSX}{{\cal D}_{\cal F} (S^D_t,X_t)}
\newcommand{\DSXA}{{\cal D}_{\cal F} (S^D_t,X_{t_k})}

\newcommand{\erre}
{
\RR
}

\newcommand{\errep}
{
\RR^+
}

\newcommand{\erren}
{
\RR^{N}
}

\newcommand{\errepm}
{
\RR^{+M}
}

\newcommand{\enne}
{
\NN
}

\newcommand{\nin}
{
\hbox{$\in\!\!\!\!\!/$}
}

\newtheorem{teorema}{Theorem}
\newtheorem{corollario}{Corollary}
\newtheorem{lemma}{Lemma}
\newtheorem{definizione}{Definition}
\newtheorem{proposizione}{Proposition}

\begin{document}
\begin{sloppypar}

\title{Throughput Optimal Scheduling Policies in Networks of Constrained
Queues}

\author{E. Leonardi}

\date{\today}


\maketitle

A shorter version is  appering on: {\it Queueing Systems: Theory and Applications } (Springer).

\begin{abstract}

This paper considers a fairly general model of constrained queuing networks that allows us to represent both  MMBP  (Markov Modulated  Bernoulli Processes) arrivals and time-varying service constraints.  We derive a set of sufficient conditions for throughput optimality of scheduling policies, which  encompass and generalize all the  results previously obtained in the field.
This leads to the definition of new classes of (non diagonal) throughput optimal scheduling policies. We  prove the stability of queues  by extending the traditional Lyapunov  drift criterion  methodology. 

\end{abstract}

\section{Introduction}

Networks of constrained queues have received significant attention from the research community 
in the last 20 years, since they provide a powerful tool for the analysis
of complex systems, such as communication, manufacturing or transportation networks.
Specifically, in the context of computer science, networks of constrained queues have been successfully applied 
to describe packet-level dynamics in wireless networks and in 
high speed Internet routers whose internal architecture is built around an
Input-Queued (IQ) switch.

In their pioneering work, Tassiulas and Ephremides~\cite{tassiulas-ephremides}, have shown that optimal throughput performance 
can be achieved in networks of constrained queues by employing a dynamic scheduling policy according to which, 
 the departure vector  maximizes the   sum of ''queue pressures'', at every time instant. The pressure of queue $q$ is defined as the difference between its own
length and the length  of the queue entered by customers leaving $q$.
The scheme proposed in~\cite{tassiulas-ephremides} is referred in the literature
as {\em max scalar}, {\em max weight}, or {\em max pressure} scheduling
policy. 
 
Since then, a large body of work has generalized the result in~\cite{tassiulas-ephremides}, mainly along four lines:
i) considering more and more general models of constrained queuing networks;~\cite{dai-lin,neely-modiano,MMBP-Tassiulas}
ii) proposing generalizations of the {\em max scalar} scheduling policy that
achieve optimal throughput;~\cite{nostro-tit,giaccone-shah,keslassy,bambos,deva-damon1,deva-damon2,linear-complexity-tassiulas}
iii) looking for simple (low computational) heuristic scheduling policies with
throughput guarantees~\cite{sarkar,walrand,srikant};
iv) attempting a characterization of delay properties of throughput optimal
scheduling policies~\cite{keslassy,nostro-jacm,neely-modiano,deva-damon1,deva-damon2}.

In particular, focusing on the second of the above mentioned aspects,
 works~\cite{nostro-tit,Erylmaz,giaccone-shah,keslassy,LPF,bambos,deva-damon1,deva-damon2,linear-complexity-tassiulas} have shown that the class of throughput optimal 
scheduling policies is significantly large. It  includes low complexity randomized scheduling algorithms~\cite{giaccone-shah,linear-complexity-tassiulas}, as well as, 
 extensions of the {\em max scalar} scheduling algorithm in which queue weights
 are possibly non linearly related to queue lengths~\cite{nostro-tit,keslassy,LPF,deva-damon1,deva-damon2}. 
 Furthermore, in networks of constrained queues with particular symmetry properties,
 scheduling policies with non diagonal weight assignments (i.e., when the weight of a queue may depend on 
 the length of other queues) have been also shown to be throughput optimal as well~\cite{LPF,bambos}. 
 
Even if the collection of results already obtained in~\cite{nostro-tit,giaccone-shah,keslassy,LPF,bambos,deva-damon1,deva-damon2,linear-complexity-tassiulas}, is rather rich, it is still far from being exhaustive. 
There are several obscure aspects that prevent full comprehension
of the structure of throughput optimal policies.
Ideally the long term final objective would be to establish a set of sufficient and {\em necessary} conditions for throughput optimality of scheduling policies.

This paper defines  a set of sufficient conditions for throughput optimality, 
 which encompasses and generalizes all previously known results. 
Our analysis is based on the application of Lyapunov functions. Our methods,
however, substantially differ from prior work because they rely on the application of more general 
Lyapunov functions, and also involve the adoption of some new
stability criteria. For the above reasons we believe that this paper provides a valuable contribution
toward a deeper understanding of the structure of throughput optimal
policies in constrained queuing networks.

This paper is organized as follows.    In Sect.~\ref{sect-queues} we introduce system assumptions and notation. 
 Previous work and paper contribution are discussed in Sect.~\ref{prev-work}. 
 Sect.~\ref{sect-lyapu} reviews
 Lyapunov drift criteria that will be invoked in the derivation of our main
 results. Sect.~\ref{sect-main} presents our main findings on throughput optimal scheduling algorithms. 
At last we conclude the paper in Sect.~\ref{sect-concl}.

\section{Preliminary definitions and notations}
\label{sect-queues}


We consider a network composed of $N$ physical queues $q_n$ with $1\le n\le N$, which may
represent, for example, either links of a wireless multi-hop network or a virtual output queues (VOQ) in a
IQ-switch architecture. 
The network is traversed by a set $\Fc$ (with $|\Fc|=F$) of different customers flows, each-one 
characterized by a given ingress/egress queue in the network $(s_f, d_f)$.

 We assume time to be slotted, and physical queues to
have infinite storage capacity. Each physical queue can potentially store
customers belonging to 
 several flows. The set of customers belonging to flow $f$ and
 enqueued in queue $q_n$ forms a virtual queue $v_{m}$.
 The whole network can be regarded as a system of $M \le F N$ discrete-time
 virtual queues represented by row vector $V$, whose $m$-th element, $1
 \le m < M$ corresponds to virtual queue $v_m$.

The routes of customer flows in the network are {\em fixed} (a priori established and time invariant). 
Without loss of generality, we assume that all customers belonging to flow $f$ and stored in queue $v_m$
will advance to the final destination following  the same simple path in
the network, which corresponds to a predetermined sequence of (virtual/physical) queues to be
traversed. We specify network routes by means of an $M \times M$ {\em routing matrix} $R=[r^{(m,p)}]$ whose element $r^{(m,p)}\in \{ 0,1\}$ indicates  whether customers
departing from virtual queue $m$ enter virtual queue $p$.~\footnote{In this
 paper the terms server and queue will be used interchangeable.} 
We remark that according to our assumptions   since all customers of a flow  residing 
in a  virtual  queue  must reach  their final destination  following the same path
 queue forking is not permitted.  Instead  queue joining (i.e., 
 multiple virtual queues feeding into one downstream virtual queue) 
 is permitted.

For any physical queue $q_n$, function $VQ(n)$ returns the set indexes
corresponding to the associated virtual queues. For every virtual queue
$v_m$, function  $PQ(m)$ returns the physical queue that corresponds to $v_m$. For any virtual queue of
index $m$, function $FL(m)$ returns the index of the corresponding customer
flow $f$. At last, for every flow $f$,  $FP(f)$ returns the ordered set of indexes of
virtual queues storing flow $f$ customers along  the associated path.

Let $X_{t}=(x^{(1)}_{t},x^{(2)}_{t},\ldots,x^{(M)}_{t})$ be the row vector whose 
 $m$-th element $x^{(m)}_{t}$, $1\le m \le M$, represents the number of
 customers (i.e., either the number of packets or bits/bytes) 
 in queue $v_m$ at time $t$. 
The evolution of the number of queued customers is described by 
$x^{(m)}_{t+1}=x^{(m)}_t+e^{(m)}_t-d^{(m)}_t$, where $e^{(m)}_t$ represents 
the number of
customers that enter virtual $v_m$ 
in time interval $(t,t+1]$, and $d^{(m)}_t$ represents the number of customers 
departures from $v_m$ it time interval $(t,t+1]$.
$E_t =(e_t^{(1)}, e_t^{(2)},\ldots, e_{t}^{(M)})$ is the vector of 
entrances in the virtual queues, and
$D_t = (d_t^{(1)}, d_t^{(2)},\ldots, d_{t}^{(M)})$ is the vector of 
departures from the virtual queues. 

With this notation, the system evolution equation can be written as:
\begin{equation}
X_{t+1}=X_t+E_t-D_t.
\label{eqeq}
\end{equation}


We represent service constraints among different servers in the network 
as follows.
At every time $t$,  the queue departure vector $D_t$ is constrained to lie within
a {\em compact} and {\em convex} region $\Dc_t$. We remark that 
region $\Dc_t$ may change over time, since it 
 is possibly controlled by a finite state discrete-time Markov chain at
 steady-state (i.e., $\Dc_t = \Dc(S^D_t)$).
 Without loss of generality we assume $\Dc(S^D_t)$ to be deterministically
 associated with  the current Markov Chain state $S^D_t$. 
 We denote by $\Sc^D$ the state space of Markov Chain $S^{D}_t$ that models possible variable environmental conditions
(such as fading conditions). Additional constraints, such as integrality
may be imposed to departure vectors $D_t$.
 However, we require that for every state $S^{D}_t$, every
vertex of $\Dc(S^{D}_t)$ represents a feasible departure vector (i.e., a vector that satisfies all constraints). 
Furthermore we assume that for any  feasible  departure vector $D\in\Dc(S^{D}_t) $ , the vector $\min(D, X_t) \in\Dc(S^{D}_t) $ (where the min is intended component-wise) 
 is feasible too.

In the particular case in which $\Dc_t = \Dc$ (i.e. $\Dc$ does not vary with
time) we say that the system of queues is subject to 
static service constraints.
We observe that this approach is very general and encompasses 
 the classical case~\cite{tassiulas-ephremides} in which service
constraints are represented by  a
contention graph.~\footnote{
Contentions graphs are typically defined as follows:
 \begin{definizione}
The {\em contention graph} $G_I({\cal V}^I, {\cal E}^I)$ is an undirected graph in which:
i) vertexes $v \in {\cal V}^I$ correspond to network (virtual) queues;
ii) an edge connects two vertexes $v$ and $v'$, if the corresponding queues can not 
simultaneously be served.
\end{definizione}}
In the latter case $\Dc$ is defined as convex hull generated by those vectors $D\in \{0,1\}^M$ that correspond 
 to independent sets of nodes over the contention graph.
$D_t\in \{0,1\}^M$, by construction, corresponds to some independent set over the contention graph, 
and therefore trivially lies in $\Dc$.
Our approach covers also the case in which $\Dc$ is determined by a rate-power
function $\mu(P_t,S^D_t)$ that 
 maps vectors of power allocations to servers
$P_t$ (under some constraint on the maximum power that can be employed) into
 vectors of service rates, for every state $S^ D_t$, as in \cite{neely-modiano}.
In this latter case $\Dc(S^D_t)$ is  the convex hull generated by 
service rate vectors that correspond to  possible extremal power allocations.

The entrance vector is the sum of two terms: 
vector $A_t =(a_t^{(1)}, a_t^{(2)},\ldots, a_{t}^{(M)})$
representing the customers arrived at the system 
from outside, and vector $J_t =(j_t^{(1)}, j_t^{(2)},\ldots, j_{t}^{(M)})$ of
recirculating customers; $j_t^{(m)}$ is the number customers  that  enter virtual queue $m$ 
in time interval $(t,t+1]$, coming from  some other virtual queue in the network.
 Note that when customers do not traverse more
that one queue (as for a switch in isolation), vector
$J_t$ is null for all $t$, and $A_t=E_t$. In this case we say that the
network is traversed by single-hop traffic. 

Let us consider the external arrival process 
$A_t =(a_t^{(1)}, a_t^{(2)},\ldots, a_{t}^{(M)})$;
in general we suppose that the sequence $A_t$ is a Markov Modulated Bernoulli
Process. We further assume the modulating Markov Chain $S^{A}_t$ to
 have a finite number of states. We denote by $\Sc^A$ its state space. At last we assume the number of arrivals at
queues  to be deterministically bounded by some constant.\footnote{We assume that of $S^D_t$ and $S^A_t$ evolve
 independently, even if 
this assumption is not strictly needed to obtain our results.} 
We denote by $ \Lambda= (\lambda^{(1)}, \lambda^{(2)},
\ldots,\lambda^{(M)})$ the average arrival vectors (arrival rates) $\EE[A_t]$.
In the specific case in which $A_t$ forms an i.i.d. sequence, we say that the traffic is i.i.d.
The workload $W_t$ provided  by customers that in time interval $[t,t+1)$ entered the system of queues 
 is given on average by $W=\EE[W_t]=\Lambda(I-R)^{-1}$,  $I$ being the identity matrix.

Note that since $J_t=D_tR$,
the system evolution equation can thus be rewritten as:
\begin{equation}
X_{t+1}=X_t+A_t-D_t(I-R)
\label{eqeq0}
\end{equation}
At last, given two vectors\footnote{In this paper $\enne$ denotes 
the set of non negative integers, $\erre$ denotes the set of real 
numbers, and $\erre^+$ denotes the set of non negative
real numbers.}, $A\in \erre^M$ and $B\in \erre^M $, we denote by $\langle A\cdot B \rangle $ the inner
(scalar) product between them $\langle A\cdot B \rangle =AB^T=\sum_{m=1}^M a^{(m)} b^{(m)}$, where
$B^T$ is the transpose of $B$;
we denote, instead, by $\|A\|$ the Euclidean norm of $A$, $\|A\|= \sqrt{\langle A\cdot A\rangle }$.

In the following we will use capital letters to denote vectors and matrices, 
lower case letters to denote scalars, calligraphic characters to denote sets. Moreover we will denote by capital
 letters, functions of multiple variables 
while by lower case letters, functions of a single variable; at last, with abuse of notation, given a vector $A$, 
we will denote by $f(A)$ the vector whose $m$-th component is $f(a^{(m)})$.

\subsection{Examples} \label{sect-examples}
As first example, we consider an input queued switch with
$N$ input ports and $N$ output ports.
The switching fabric is assumed to be non-blocking and memoryless.
Fixed size packets are stored  at input ports. Thus, one physical queue 
corresponds to every input port. 
Each input port maintains a separate virtual queue for each output port. Therefore,
the considered switch  can be modeled as a system comprising $M=N^2$ virtual 
queues. Let $v_m,\ m=iN+j$ be the virtual queue at input $i$ 
storing packets  directed to output $j$, with $i,j=0,1,2,\ldots,N-1$.

At each  time slot, the switch scheduler selects packets to be 
transferred from input ports to output ports. The set of packets to be
transferred during an internal time slot must
satisfy two constraints: 
i) at most one packet can be transferred from  each
input port, and
ii) at most one packet can be transferred toward each output.
Service constraints can be formalized as:
\[
\sum_ {m\in VQ_I(i)}  d_t^{(m)}  \le 1  \qquad \sum_ {m\in VQ_O(j)}   d_t^{(m)}\le 1 \qquad  \forall i,j
\]
where  $VQ_I(i)$ denotes  the set of  indexes 
associated to VOQs storing packets at input  $i$; and 
 $VQ_O(j)$,  the set of indexes of   VOQs storing packets directed  to output $j$.

As second example we consider a ad-hoc network    with $N$  nodes.
Every node is  provided with a single transmitter and   
maintains a  per  destination   virtual queuing structure. 
Thus,  at  node $i$    packets destined to  node $j$ are enqueued in a virtual queue $v_m$  
with $m=iN+j$. The system of queues can be modeled as a system of $M=N^2$  virtual queues. 
Packet routes are assumed fixed;
all packets at node $i$ destined to node $j$ follow the same route to their destination. 

Service constraints  come from the fact that; 
1) two virtual queues  residing in the same node (i.e., insisting on the same physical queue)
 can not be activated simultaneously because they conflict
for the same physical transmitter.
2)  some pairs of virtual queues residing in different nodes can not be activated (served) 
simultaneously because of  mutual interference on  the receivers.
Contention graph $G_I({\cal V}^I, {\cal E}^I)$
fully specifies services constraints.

\subsection{Stability Definitions}
\label{sect-stabdefin}

Several stability criteria for constrained queuing networks have being defined 
in the technical literature:
\begin{definizione}
A system of queues is {\em rate-stable} if
\[
\lim_{t\to \infty} \frac{X_t}{t}=
\lim_{t\to \infty} \frac{1}{t}\sum_{\tau=0}^{t-1} (E_\tau-D_\tau) = 0
\quad \mbox{with~probability~1}.
\]
\label{def:100thru}
\end{definizione}


\begin{definizione}
A system of queues is {\em weakly stable\/} if, for every $\epsilon >0$,
there exists a $b>0$ such that:
$$
\lim_{t\to \infty} \Prob\{\|X_t\|>b\}<\epsilon
$$
where $\Prob\{\Ec\}$ denotes the 
probability of event $\Ec$.
\label{def:weakly}
\end{definizione}
\begin{definizione}
A system of queues is {\em strongly stable\/} 
if
$$
\limsup_{t\to \infty} \EE[\|X_t\|] <\infty
$$
\label{def:strong}
\end{definizione}
Note that strong stability entails weak stability, and that weak stability
entails rate-stability. Indeed, rate stability allows queue
lengths to indefinitely grow with sub-linear rate, while the weak stability
 entails that queues are finite with probability 1. This however does
not guarantee that the average delay experienced by customers is bounded.
Strong stability entails, in addition, the boundedness of average 
customer delays.

Strong-stability concept can be generalized as follows~\footnote{$C^k[\erre \to
\erre]$ denotes the class of real valued functions that are $k$-th times
continuously differentiable.
Furthermore given a sufficiently smooth function $g(x)$: $\erre \to \erre$ we denote by $g'(x)$
its first derivative, with $g''(x)$ its second derivative, and with
$g^{(h)}(x)$ its $h$-th derivative.} :
\begin{definizione}
Given a non-negative continuous function $F(X)\in C[\erre^{M}\to \errep]$, 
 with $\lim_{\|X\|\to \infty}F(X)=\infty$;
 a system of queues is $F(X)$-{\em stable\/}
if
$$
\lim_{t\rightarrow \infty} \sup \EE[F(X_t)] <\infty
$$
\label{def:fstability}
\end{definizione}
Note that $F(X)$-{\em stability \/} property becomes stricter
by selecting functions $F(X)$ that increase faster to $\infty$, 
for large $\|X\|$. In other words $F(X)$-{\em stability\/} entails $G(X)$-{\em
 stability\/} for any other function $G(X)$ such that \footnote{Given two functions $f(x)\ge\! 0$ and $g(x)\ge\! 0$: \mbox{$f(x)\!=o(g(x))$} means
$\lim_{x \to \infty}{f(x)}/{g(x)}=0$; $f(x)=O(g(x))$ means $\limsup_{x \to
\infty}{f(x)}/{g(x)}=c<\infty$. 
} $G(X)=O(F(X))$ as~\footnote{For any function $F: \errepm\to \erre$  we use  $\lim_{\|X\| \to \infty}F(X)=l$ with   $l\in \erre \cup \{\infty\}$
 as shorthand  notation  to mean that 
  $\lim_{\|\alpha \| \to \infty}F(\alpha X_0)=l$   
 for any $X_0 \in \errepm$ with  $\|X_0\|=1$}  $\|X\|\to \infty$. 
In the following we will make extensive use of the $F(X)$-{\em stability \/}
criterion.

\subsection{Capacity Region} \label{sect-capreg}
Given a scheduling policy $\pi$, 
the stability region of a network of queues is the set of average arrival
vectors (arrival rates) $\Lambda$ in correspondence of  which the system is stable (under one of the above criteria). 
Arrival rate $\Lambda$ is said to be admissible when it lies in the
stability region for some scheduling policy $\pi'$. The capacity region of 
the network is the set of all admissible arrival rates i.e. the set of vectors for  which there exists some scheduling policy that makes the system of queues stable.
With  abuse of language we say that    arrival process is {\em admissible} if  its rate is admissible. 

Under the rate stability criterion, the capacity region of the system $\Cc_{\rm rate}$, 
 is given by the set of $\Lambda$:



\begin{equation}
\Cc_{\rm rate}=\left\{ \Lambda : W=\Lambda(I-R)^{-1} = \sum_{S^D \in \Sc^D}\pi_{S^D} D(S^D) \right\} \mbox{ with } D(S^D)\in \Dc(S^D), \forall S^D\in \Sc^D 
\label{capacity-region}
\end{equation}
where $\pi_{S^D}$ is the steady state probability associated with states $S^D \in \Sc^D$ of the DTMC governing service constraints, and $D(S^D)$ is an arbitrary vector lying  in  $\Dc(S^D)$ ~\cite{neely-modiano,tassiulas-ephremides}.
Observe that $\Cc_{\rm rate}$ is a compact (closed and bounded) set in $R^{+N}$.
Under either the weak and strong stability criterion,
the capacity region $\Cc_{\rm weak}=\Cc_{\rm strong}$  corresponds 
to the interior of $\Cc_{\rm rate}$, i.e. to the set of average arrival vectors $\Lambda$,  whose corresponding workloads $W$ that can be written in the form:
 $W = \sum_{S^D \in \Sc^D}\pi_S D(S^D) $, with $ D(S^D)$ lying in the interior of $\Dc(S^D)$.

\section{Previous Work and Paper Contribution}
\label{prev-work}
In their seminal work, Tassiulas and Ephremides \cite{tassiulas-ephremides} have shown that 
under i.i.d. arrival processes and static service constraints,
optimal throughput can be achieved by employing {\em max
scalar} scheduling policy $\pi_{\max}$, according to which 
at every time slot $t$, the departure vector, satisfies:
\[
D^{\max}_t =\argmax_{\DX}
\langle X_t(I-R)^T\cdot D \rangle
\]
where $\DX$  represents the set of feasible departure vectors  $D\in \Dc$ satisfying $D\le X_t$.
 
More precisely $\pi_{\max}$ guarantees the network of queues
to be weakly stable within the capacity region. Observe that the queue length
vector $X_t$ has to be interpreted as a vector of {\em weights} associated
to queues, while $X_t(I-R)^T$ is the corresponding vector of {\em pressures}
that take into account the effect of customers recirculation (for networks of
queues supporting single-hop traffic, pressures coincide with weights).

The result in~\cite{tassiulas-ephremides} has been extended in several
respects.
First, the stability properties of the {\em max scalar} policy have been strengthened (strong
stability has been proved) and
extended under more general non i.i.d. traffic and dynamic
service constraints~\cite{dai-lin,neely-modiano}. 

Second, the class of throughput optimal schedulers has been extended,
including {\em max scalar} policies that employ non linear queue weights. Under i.i.d. arrival processes and static service constraints, 
scheduling policies according to which the vector of departures satisfies:
\[
D^{g}_t =\argmax_{\DX } \langle g(X)(I-R)^T \cdot D \rangle 
\]

where $g(x) \in C^1[\errep \to \erre]$ is a non negative function satisfying: $g(0)=0$ and 
$\lim_{t\to \infty}\frac{g'(x)}{g(x)}=0$, have been shown to be throughput
optimal \cite{nostro-tit,Erylmaz,keslassy,deva-damon1,deva-damon2,shah-sig}.
Particularly relevant are the cases in which 
$g(X)=X^\alpha$ for $\alpha>0$. Despite the fact that strong stability has been analytically proved for $\alpha<1$  very recently \cite{shah-sig}, it is a longstanding conjecture \cite{keslassy,deva-damon1,deva-damon2} that optimal delay properties are achieved when $\alpha\to 0$. In \cite{deva-damon1,deva-damon2} this conjecture has been supported 
by some analytical evidence.

Non-diagonal {\em max scalar}  policies achieving optimal throughput performance, 
 have been  have recently  identified  in~\cite{meyn,bambos}.
In~\cite{bambos}  {\em Projective Cone Schedulers } PCS, 
 a new class of scheduling policies has
been shown to be throughput optimal (under the rate stability criterion) in networks transporting single-hop traffic.
According to PCS the departure vector at every time $t$ satisfies:
\begin{equation} 
D^{PCS}_t =\argmax_{\DX} \langle XQ \cdot D\rangle 
\label{cone}
\end{equation}
where $Q$ is a positive definite symmetric matrix with null or negative out of diagonal elements.
Observe that according to PCS, contrarily to all previously mentioned schemes, 
 weight associated with queue $v^{(m)}$ may depend on the length of other queues. In this case we say that the scheduling policy employs non diagonal
weights. Moreover, we wish to mention that other examples 
of policies employing non diagonal weights have been earlier 
shown to achieve throughput optimality in constrained queuing networks with particular structures,
such as those corresponding to IQ switches (see for example LPF for IQ switches~\cite{gupta-sig,LPF}).

{A different result has been obtained   in~\cite{meyn}.   
For a general network with static service constraints, given  a function  $G(X)$,  $G \in C^1[\erre^{+M}\to \errep]$,  the scheduling policy:
\begin{equation} 
D^{\gmax}_t =\argmax_{\DX } \langle \nabla G(\hat{X}_t)(I-R)^T\cdot D\rangle,
\label{meyn}
\end{equation}
with $\hat{X}_t=X_t+ \theta [e^{-X_t/\theta}-1]$ for $\theta\ge 0$,
has been proven to be throughput optimal, provided that 
$ G(X)$   is monotonic,  i.e.  $\nabla G(X) \in \errepm $  for any  $X \in \errepm$;
 $\|\nabla G(X)\|$ is Lipschitz continuous;  $\|\nabla G(X)\| \to \infty $ as $\|X \| \to \infty$; $\frac{\partial G(\hat{X})}{ \partial x_k}=0$ when  $x_k=0$.
Observe, however,   previous requirements such as  monotonicity,  severely reduce the domain  of  applicability of the result in~\cite{meyn}.
 For example, functions $G(X)$ associated to  non trivial {\em Projective Cone Scheduler}  
(with negative out of diagonal elements)  are not monotonic. 
 Our analysis  generalizes~\cite{meyn} making a further significant step 
in the direction  of  the  identification of  the most  general set of
conditions for $G(X)$,  which guarantee throughput optimality for   the 
associated {\em max-scalar} policy.

}

Scheduling policies with memory~\cite{giaccone-shah,modiano-shah,linear-complexity-tassiulas}  represent a further
example of throughput optimal schemes for networks with static service constraints. 
 The schemes proposed in~\cite{giaccone-shah,modiano-shah,linear-complexity-tassiulas}
 are based on the idea of generating  an admissible 
candidate departure vector $D^c_t$ at every slot, according to some
simple rule; then the departure vector $D^{\mems}_t$ is
 selected between  $D^c_t$ and $D^{\mems}_{t-1}$
 by maximizing the  associated aggregate pressure $D^{\mems}_t=\argmax \{\langle X \cdot D^c_{t}\rangle , \langle X \cdot D^{\mems}_{t-1}\rangle \}$.
 It has been shown that such schemes achieve optimal throughput (i.e., strong stability) 
under admissible i.i.d. arrival processes and static constraint conditions, 
provided that at every slot it can be guaranteed $D^c_t=\argmax_{\DX} \langle X \cdot D \rangle $ 
with a probability that is not small  than $\delta>0$. Notice that the above condition
is satisfied when $D^c_t$ is uniformly  selected among vectors in $\DX$.


This paper provides several contributions with respect to previous work:
i) Theorem \ref{max-scalar-theo-sh} and \ref{max-scalar-theo-sh-strong}
 significantly extend of  the class of throughput optimal {\em max scalar} like policies policies exploiting non linear and non diagonal weights.
{ In particular with respect to \cite{meyn} , Theorems \ref{max-scalar-theo-sh} and \ref{max-scalar-theo-sh-strong} do not require  $G(X)$ to be quadratic and monotonic.}
 Moreover, throughput optimality is proven under a general model of constrained queuing networks possibly subject to dynamic service constraints and non i.i.d. arrivals.
ii) Theorems \ref{memory} and \ref{memory-strong}  
generalize  the class of throughput optimal scheduling algorithms with memory,
 applying, for the first time to the best of our knowledge,
 the concept of schedulers with memory to network of constrained queues 
subject to dynamic service constraints. 
iii) We strengthen the
above results, showing that every polynomial moment of the queue lengths
remains finite under any of the above schemes,
as long as the average arrival vector lies within the capacity region.
iv) At last, from a methodological point of view, 
we  introduce new Foster-Lyapunov drift conditions for $F(X)$-stability (reported in
Sect.~\ref{sect-lyapu}), extending in such a way   previous drift 
arguments.

\section{Markov State and Lyapunov Stability Criteria}
\label{sect-lyapu}
Under previous assumptions, the process describing the evolution of the system of
queues is an irreducible Discrete-Time Markov Chain (DTMC), whose
state vector at time $t$, $Y_t=(X_t, S_t)$, is the combination of vector $X_t$ and  vector $S_t$
that represents the memory of the system in
the case in which arrivals are not i.i.d. and/or service constraints are dynamic.

Let $\Hc$ be the state space of the DTMC, obtained as Cartesian product of the
state space~\footnote{$\enne$ denotes the set of non negative integers.}
 $\Xc\subseteq \enne^{M}$ induced by the queue lengths vector $X_t$
and the state space $\Sc=\Sc^A \times \Sc^D \subset \enne^{K} $ induced by $S_t$, we further assume $\Sc$ to be a 
{\em finite\/} state space. Note that $\Hc\subset \enne^{+H}$ with $H=M+K$.

From Definition~\ref{def:weakly}, we can immediately see that DTMC $Y_t$ is
positive recurrent, if and only if the system of queues is weakly stable (we recall that the DTMC modelling the system is assumed to be irreducible).

The following general criterion for the (weak)
stability of systems is
therefore useful in the design of scheduling algorithms.
This theorem is a
straightforward extension of Foster's Criterion; see
\cite{kushner,tassiulas-ephremides}.

\begin{teorema}
\label{foster}
Given a system of queues  described by a DTMC
with state vector $Y_t=(X_t,S_t) \in \enne^{H}$, whose 
 state space $\Hc$ is  the Cartesian product of the denumerable state space  $\Xc\subseteq \enne^M$ (with
 $X_t\in \Xc$),  and a finite state space
$\Sc\in \enne^{K}$ (with $S_t\in \Sc$); 
if a lower bounded continuous function $\Lc(Y)$, called Lyapunov function,
$\Lc: \erre^{+H} \rightarrow \erre$ can be found such that:
\begin{equation}
\EE[ \Lc(Y_{t+1}) \mid Y_{t} ] < \Lc(Y_{t})+v_0 
\label{lyapunov0}
\end{equation}
for some $v_0<\infty$, and 
\begin{equation}
\EE[\Lc(Y_{t+1})- \Lc(Y_{t}) \mid Y_{t} ] < -\epsilon \quad \forall Y_t:\|X_{t}\|>b,
\label{liapunov1}
\end{equation}
for some $\epsilon \in \erre^+$ and $b \in \erre^+$;
then the DTMC is positive recurrent and the system of queues is
weakly stable.
\end{teorema}
{\bf Remark:} observe that for every $Y_t:\|X_{t}\|>b$, the satisfaction of 
(\ref{lyapunov0}) immediately follows from (\ref{liapunov1}) (with
$v_0=0$). Therefore, it is sufficient to verify (\ref{lyapunov0}) 
for $Y_t:\|X_{t}\|<b$ and (\ref{liapunov1}) to apply the above Theorem.
The following result provides a criterion for strong stability.

\begin{teorema}
 Under the same assumptions of Theorem \ref{foster}, if $\Lc(Y)$, 
additionally satisfies:
\begin{equation}
\EE[ \Lc(Y_{t+1})- \Lc(Y_{t}) \mid Y_{t} ] < -\epsilon \|X_t\| \quad \forall
Y_t:\|X_{t}\|>b,
\label{outside-foster}
\end{equation}
 for some $\epsilon \in \erre^+$ and $b \in \erre^+$; then the system of queues is {\em strongly}-stable.
\label{th2}
\end{teorema}
Previous criteria can be also applied to establish the stability of
a DTMC $Y_{t_k}$, obtained by sampling $Y_{t}$ in correspondence of an
opportunely defined  sequence of time instants.
In particular we are interested in the case in which $t_k \in
\enne^+$ form a sequence of stopping times:
\begin{definizione}
A sequence of  random time instants $t_k \in \enne^+$ 
is a sequence of {\em non-defective}
regeneration instants (or stopping times)  for the evolution of a system of queues iff:
i)  for any $k$,   the event $\{t_k = t\}$  belongs to the $\sigma$-algebra  defined by past trajectories $[Y_1, Y_2, Y_3, \cdots, Y_t]$. 
ii)  variables $z_k=t_{k+1}-t_k$ are identically distributed and satisfy:  $\EE[(z_k)^h]<\infty$, for any $h\in \enne^+$.
\end{definizione}
From the strong Markov property~\cite{tweedie-meyn}  immediately  follows that  
the evolution of Markov Chain $Y_t$ after $t_k$ 
is conditionally independent of the evolution 
of the system before $t_k$, given the state $Y(t_k)$,  provided that $t_k$ is a stopping time. 
We remark, instead,  that  the above conditional independence property  does not hold if $t_k$ is a generic random time.

From the strong stability of $Y_{t_k}$ it is possible to infer
strong stability of the original system:
\begin{teorema}
 Under the same assumptions of Theorem \ref{foster},
and the additional assumption that both arrival vectors, $A_t$, and 
 departure vectors, $D_t$, are bounded in norm, 
 if a lower bounded continuous Lyapunov function $\Lc(Y)$,
$V: \erre^{+H} \rightarrow \erre$ can be found such that, 
for an opportunely defined {\em non-defective} sequence of
regeneration instants $\{t_{k}\}$:
\begin{equation}
\EE[ \Lc(Y_{t_{k+1}}) \mid Y_{t_k} ] < \Lc(Y_{t_k})+v_0 
\label{inside-sampled}
\end{equation}
for some $v_0<\infty$,
and
\begin{equation}
\EE[\Lc(Y_{t_{k+1}})- \Lc(Y_{t_k}) \mid Y_{t_k} ] < -\epsilon \|X_{t_k}\| \quad \forall Y_{t_k}:\|X_{t_k}\|>b
\label{outside-sampled}
\end{equation}
for some $\epsilon \in \erre^+$ and $b \in \erre^+$; then the system of queues is {\em strongly}-stable.
\label{th3}
\end{teorema}
A brief proof of this statement is in Appendix \ref{app-proofs3}.

Lyapunov drift arguments can be extended to obtain the following criterion for 
$F(X)$-stability:

\begin{teorema}
 Under the same assumptions of Theorem \ref{foster},
if it can be found a lower bounded continuous Lyapunov function $\Lc(Y)$, 
$\Lc: \erre^{+H} \rightarrow \erre$ 
 satisfying the following two conditions:
\begin{equation}
\EE[ \Lc(Y_{t+1}) \mid Y_{t} ] < \Lc(Y_{t})+v_0 
\label{inside-F}
\end{equation}
for some $v_0<\infty$, and 
\begin{equation}
\EE[ \Lc(Y_{t+1})- \Lc(Y_{t}) \mid Y_{t} ] < -\epsilon F(X_t) \quad \forall Y_t:\|X_{t}\|>b
\label{outside-F}
\end{equation}
for some $\epsilon \in \erre^+$, $b \in \erre^+$,  being $F(X):\erre^{+M}\to \errep$  continuous,  with
$\lim_{\|X\|\to \infty}F(X)=\infty$;
then the system of queues is $F(X)$-stable.
\label{th4}
\end{teorema}
The proof is reported in appendix.

At last, using similar arguments as in Theorem \ref{th3},
 we can easily derive the following result:
 \begin{corollario}
 Under the same assumptions of Theorem \ref{foster},
and the additional assumption that that both arrival vectors $A_t$ and 
departure vectors $D_t$ are bounded in norm,
 if a lower bounded continuous Lyapunov function $\Lc(Y)$, 
$\Lc: \erre^{+H} \rightarrow \erre$ can be found such that:

\begin{equation}
\EE[ \Lc(Y_{t_{k+1}}) \mid Y_{t_k} ] < \Lc(Y_{t_k})+v_0, 
\label{inside-sampledF}
\end{equation}
for an opportunely defined sequence $\{t_{k}\}$ of
{\em non-defective} regeneration times and for some $v_0<\infty$; 

\begin{equation}
\EE[\Lc(Y_{t_{k+1}})- \Lc(Y_{t_k}) \mid Y_{t_k} ] < -\epsilon F(X_{t_k}) \quad \forall Y_{t_k}:\|X_{t_k}\|>b
\label{outside-sampledF}
\end{equation}
for some $\epsilon \in \erre^+$ and $b \in \errep$: 
being $F(X):\erre^{M}\to \errep$, a  continuous function with
$\lim_{\|X\|\to \infty}F(X)=\infty$;
then the system of queues is $F(X)$-stable.
\label{th3a}
\end{corollario}

\section{Main Results}
\label{sect-main}

In this section we introduce the class of scheduling policies that 
achieve optimal throughput performance. 
To improve the readability of the section, all proofs have been moved to Appendix \ref{app-proofs5}.

\begin{definizione}
Given any function
 $G(X)$, $G \in C^1[\erre^{+M}\to \erre]$, 
we define as $\nabla G(X)$-{\em max scalar\/}, the scheduling policy
$\pi_{\gmax}$ that 
selects the departure vector according to:
\begin{equation}
D_t^{\gmax}=\argmax_{\DSX}
\langle \nabla G(X_t)(I-R)^T \cdot D \rangle , 
\label{max-scalar-rete}
\end{equation}
where $\DSX$ represents  the set of feasible departing vectors at time $t$  (i.e.,
 $D\in \Dc(S^D_t)$ and $D\le X_t$, $D$ feasible).   

In other words  $D^{\gmax}_t$ is the feasible vector of departing customers in $\Dc(S^D_t)$  satisfying  $D^{\gmax}_t\le X_t$ that  maximizes
the inner product between the departure vector itself, and the gradient of $G(X)$ evaluated at $X_t$,
 ($\nabla G(X)\mid_{X=X_t}$, denoted for short by $\nabla G(X_t)$), 
 multiplied by the transpose of matrix $(I-R)$. 

\end{definizione}
Note that $\nabla G(X_t)(I-R)^T$ can be interpreted as the vector of
pressures associated with the weight vector $\nabla G(X_t)$.
Furthermore, observe that since $\langle \nabla G(X_t)(I-R)^T \cdot D\rangle
=\langle \nabla G(X_t)\cdot D(I-R) \rangle $,
the $\nabla G(X)$-{\em max scalar\/} can be defined as well as scheduling policy according to which: 
\begin{equation}
D_t^{\gmax}
=\argmax_{\DSX}
\langle \nabla G(X_t)\cdot D(I-R)\rangle. 
\label{max-scalar-rete2}
\end{equation}
At last, in the relevant case in which the network is traversed by 
single-hop traffic, i.e. when $R=0$,
$D_t^{\gmax}$ satisfies: 
\begin{equation}
D_t^{\gmax}=\argmax_{\DSX}
\langle \nabla G(X_t) \cdot D\rangle. 
\label{max-scalar-single}
\end{equation}

The following two theorems provide conditions for throughput optimality of 
 $\nabla G(X)$-{\em
max scalar\/} scheduling policies. 
We recall that an arrival process  is said admissible if its 
associated  average workload $W=\Lambda(I-R)^{-1} $ lyes in the convex hull of the  of the
 feasible departure vectors, i.e., departure vectors that satisfy service constraints.
We denote with $H_G(X)$ the Hessian of $G()$ at $X$  

\begin{teorema} 
\label{max-scalar-theo-sh}
The network of queues is $\|\nabla G(X)\|$-stable under i.i.d.
admissible arrival processes  and static service constraints, whenever a $\nabla G(X)$-{\em
max scalar\/} scheduling policy is employed, provided that 
 $G(X)$ is in $ C^2[\erre^{+M}\to \erre]$ and satisfies the following technical conditions:

\begin{enumerate}
\item $G(X)$ grows to infinity faster than ${\|X\|}$ when $X$ grows to
infinity,~\footnote{We recall that for  any function $F: \errepm\to \erre$   we use  $\lim_{\|X\| \to \infty}F(X)=l$ with   $l\in \erre \cup \{\infty\}$
 as shorthand  notation  to mean that 
  $\lim_{\|\alpha \| \to \infty}F(\alpha X_0)=l$   
 for any $X_0 \in \errepm$ with  $\|X_0\|=1$} 
 i.e.: 
\begin{equation}
\lim_{\|X\| \to \infty} \frac{G(X)}{\|X\|}=\infty;
\label{asympG}
\end{equation}
\item $G(X)$ exhibits a sub-exponential behavior for large $X$; i.e, 
\begin{equation}
\lim_{\|X\| \to \infty} \frac{G(X+Y)}{G(X)}=1, \quad 
\lim_{\|X\| \to \infty}\displaystyle \frac {\langle \nabla G(X+Y) \cdot Z \rangle}{\langle \nabla G(X) \cdot Z\rangle }=1,\quad
\lim_{\|X\| \to \infty}\displaystyle \frac { Z H_G(X+Y) Z^T }{ Z H_G(X) Z^T }=1,
\label{asymptderadd}
\end{equation}
for arbitrary bounded vectors $Y$, $Z$;

\item the following conditions on the orientation of $\nabla G(X) $
are met:
\begin{equation}
\langle \nabla G(X)(I-R)^T \cdot D\rangle \le 0\;\;\;
\forall D \ge 0 \mbox{ s.t. } \langle X \cdot D \rangle =0.
\label{negder}
\end{equation}
and
\begin{equation}
\lim_{\|X\| \to \infty} \frac{ \langle \nabla G(X)(I-R)^T \cdot D\rangle}{\|\nabla G(X)\|} > 0 
\mbox{  for some }D\ge 0
\label{posorien}
\end{equation}

\end{enumerate}


\end{teorema}

Stability properties of $ \nabla G(X)$-{\em max scalar\/} scheduling
policies can be extended to
more general  Markov Modulated Bernoulli Process (MMBP) arrival
processes and dynamic service constants, when $G(X)$ satisfies slightly
less general conditions:

\begin{teorema}
\label{max-scalar-theo-sh-strong}
The network of queues is $\|\nabla G(X) \|$-stable under admissible
MMBP arrival processes and general service constraints whenever a
 $ \nabla G(X)$-{\em max scalar\/} scheduling policy is employed, 
provided that $G(X)$ is in $C^\infty[\erre^{+M}\to \erre]$ and meets
the following two conditions: 
\begin{enumerate} 
\item \begin{equation}
\limsup_{\|X\| \to \infty} \|{(\partial^{h_0} G) (X)}\| <\infty \quad \mbox{for some } 
h_0\in \enne;
\label{asymptder-strong2} 
\end{equation}

\item
\begin{equation}
\lim_{\|X\| \to \infty} \displaystyle \left| \left|\frac {(\partial^{h+1}G) (X)}
{ (\partial^{h}) G(X) } \right|\right|=0 \quad \forall h<h_0;
\label{asymptder-strong} 
\end{equation}
\end{enumerate}
in addition to (\ref{asympG}), (\ref{negder}) and (\ref{posorien}). 
\end{teorema} 
Observe that conditions (\ref{asymptder-strong2}) and (\ref{asymptder-strong}), which entail  
  (\ref{asymptderadd}), express 
the fact that the dominant behavior of $G(X)$ for $\| X \|\to \infty$ is 
polynomial.

When $G(X)$ satisfies the 
technical conditions specified by Theorem~\ref{max-scalar-theo-sh}, we say that it 
is a weak-potential for the system of queues; we, instead, say that it 
is a strong-potential for the system of queues,
 when $G(X)$ satisfies the additional
technical conditions specified by Theorem \ref{max-scalar-theo-sh-strong}.
We recall that the proofs of Theorems \ref{max-scalar-theo-sh} and \ref{max-scalar-theo-sh-strong} are in 
Appendix \ref{app-proofs5}.

Note that according to Theorems \ref{max-scalar-theo-sh} and
\ref{max-scalar-theo-sh-strong},   $\|\nabla G(X)\|$-stability  has been proved 
for  $\nabla G(X)$-{\em max scalar} policies in non overloaded conditions.
$\|\nabla G(X)\|$-stability  may become weak, especially when $ \nabla
G(X)$ increases slowly to infinity for $\|X\|\to \infty$.
For example if $G(X)=\frac{1}{1+\alpha}\sum_{m} (x^{(m)})^{1+\alpha}$ for $\alpha<1$ (i.e.
$\nabla G(X)=X^{\alpha}$), strong stability of the network of queues is not
guaranteed by the above mentioned Theorems. 
Following Corollary allows us to strengthen Theorem \ref{max-scalar-theo-sh} and Theorem
\ref{max-scalar-theo-sh-strong}, showing that $\nabla G(X)$-{\em
max scalar\/} policies associated with weak/strong potentials guarantee that 
every polynomial moment of queue-lengths remains finite within the capacity region:

\begin{corollario}
Consider a weak potential function $G(X)$;
 the network of queues is $\|X\|^h$-stable, for any $h\in \enne$ (i.e., every polynomial 
moment of the queue lengths is finite), under admissible
i.i.d. arrival processes and static service constraints, provided that the associated $ \nabla
G(X)$-{\em max scalar\/} scheduling policy is employed.
When, instead, $G(X)$ is a strong potential function, $\|X\|^h$-stability 
 can be proved  for any $h\in \enne$, under MMBP arrival processes and dynamic service constraints.
\label{strong-stab}
\end{corollario}
Again, we recall that the proof of the Corollary is in Appendix~\ref{app-proofs5}.

{\bf Remark:} The class of scheduling policies that satisfy the assumptions of
Theorem~\ref{max-scalar-theo-sh} (or Theorem~\ref{max-scalar-theo-sh-strong})
is fairly large and comprises  the following three subclasses of optimal policies,
 as particular cases. Indeed note that:

\begin{enumerate}
\item Any function $G(X)$ in the form: $ G(X)= \sum_{m}=g(x^{(m)})$, where 
 $g(x)$ a function in $C^2[\errep\to \erre]$ 
 with a super-linear and sub-exponential asymptotic behavior, (i.e. $g(x)$
 such that:  
 $\lim_{x\to
 \infty} \frac{g(x)}{x}= \infty$,  $\lim_{x\to
 \infty} \frac{g(x+1)}{g(x)}= 1$,  
 and $\lim_{x\to \infty} \frac{g`(x)}{g(x)}=\frac{g''(x)}{g'(x)}= 0$), 
 and with the first derivative null in the origin ($g'(0)=0$),  is a weak
 potential.
Furthermore If $g(x)$ is in $C^\infty[\errep\to \erre]$ and has a polynomial
asymptotic behavior for large $x$, (i.e., $\limsup_{x\to 
\infty} g^{(h_0)}<\infty$ for some $h_0 \in \enne$, and
 $\lim_{x\to \infty} \frac{g^{(h+1)}(x)}{g^{(h)}(x)}=0$ $\forall h< h_0$), then $G(X)$ is a strong
potential. The associated $\nabla G(X)$-{\em max scalar} policy, according to which $D= \argmax \langle h(X) \cdot D\rangle $ with $h(x)=g'(x)$ 
 achieves $\|X\|^h$-stability for any $h$. 
 With abuse of language when $g(X)$ satisfies the above conditions, we say that it is a
 weak (strong) scalar potential.  For this subclass of scheduling policies,
we extend 
findings in~\cite{nostro-tit,keslassy,deva-damon1,deva-damon2,shah-sig}, 
since we prove  a stronger form of stability (the finiteness of every polynomial moment) 
under a more general network model with  possibly correlated
arrivals and dynamic service constraints.
As a particular case, if we select $f(x)=\frac{x^{\alpha+1}}{(\alpha+1)} $ 
we
obtain $D_t= \argmax \langle X^{\alpha} \cdot D \rangle $. By choosing, instead $f(x)=(x+1)(\log (x+1) - 1)$
we can prove stability properties of the scheduling policy according to
which $D_t= \argmax \langle \log(1+ X) \cdot D \rangle $. 

\item Choosing $ G(X)= \langle g(X)Q\cdot g(X)\rangle $
we obtain another subclass of functions
satisfying the assumptions of Theorem~\ref{max-scalar-theo-sh-strong} for networks 
transporting single-hop traffic, provided that 
$Q$ is a positive definite symmetric matrix with
non positive off-diagonal elements, and $g(x)$ is $C^\infty[\errep\to \erre]$, increasing, null in the origin (i.e., $g(0)=0$)
 with polynomial asymptotic behavior for large $x$, 
 (i.e., $\limsup_{x\to 
\infty} g^{(h_0)}<\infty$ for some $h_0 \in \enne$, and
 $\lim_{x\to \infty} \frac{g^{(h+1)}(x)}{g^{(h)}(x)}=0$ $\forall h<h_0$)
 and such that $\lim_{x\to \infty} \frac{g(x)}{\sqrt{x}}= \infty$.

This class of functions extends the class of scheduling policies 
proposed in \cite{bambos}, for  which 
$D= \argmax \langle XQ \cdot D \rangle $ (obtained when $g(x)=x$).  
{  Once again,  we wish to emphasize  this class {\em $G(X)$-max scalar} policies 
    is not covered by~\cite{meyn} (even for  $g(x)=x$),  since $G(X)$ is  not monotonic, 
as effect of the negative out of diagonal elements of $Q$.}

\item For networks transporting single-hop traffic,
 every $G(X)$ in the form $G(X)= f(X)Pf(X)^T= \langle f(X)P \cdot f(X)\rangle$ can be easily shown to be a strong potential, provided that:
 i) $P$ is a strictly  positive definite
 symmetric matrix, ii) $f(x)$ is given by: 
 \[
f(x)=x+\theta(e^{-x/\theta}-1).
\]
with  $\theta>0$.
 In particular, the above function satisfies
(\ref{negder}) since $f(0)=0$ and $f'(x)\mid_{x=0}=0$.
 This class of policies corresponds to the class of policies defined in \cite{meyn} when we add the extra constraint that every off-diagonal 
element of $P$ is non negative so  to 
guarantee monotonicity. In addition it   generalizes  LPF
policy~\cite{gupta-sig,LPF} defined for input queued switch architectures.
{To establish a clearer relationship between  LPF and the  class {\em $\nabla  G(X)$-max scalar} 
 policies with $G(X)= \langle f(X)P \cdot f(X)\rangle$,  we  focus on   networks of queues  with static service
 constraints. Without loss of generality, we assume service constraints among  virtual queues  
to be represented  by a contention graph.       For any  virtual queue $v_m$ 
 we can define  ${\cal I}_m$,   the set of  virtual queues  that  are  conflicting  with $v_m$.  
We conventionally assume $v_m \in {\cal I}_m$.
Then   taking matrix $P$,    such that;   its element $p_{m,m'}=1$  if $m' \in {\cal I}_m$ (and  by construction   $m \in {\cal I}_m'$) and   $p_{m,m'}=0$  otherwise; we obtain  a  {\em max scalar} scheduling policy  whose associated queue weights satisfy:
 $$w_m=\nabla G(X)\mid_m  = (1- e^{-x_m/\theta})\sum_{m': v_{m'} \in {\cal  I}_m} x_{m'} +\theta (e^{-x_{m'}/\theta}  -1).$$

Now if we consider an IQ  switch  architecture queue architecture, for any VOQ $v_m$,
 ${\cal I}_m$   is,  by construction, composed of all the virtual queues  residing on the same 
input port or directed to the same output port of the VOQ $v_m$.   
Thus, {\em $\nabla  G(X)$-max scalar} scheduling policy  associated to $G(X)= \langle f(X)P \cdot f(X)\rangle$
 degenerates into  a   
 LPF policy, with slightly modified queue weights.  
}

\end{enumerate}
The above three sub-classes of optimal policies are not at all exhaustive.
For example, 
functions in the form $ G(X)= \langle g(X)P\cdot g(X)\rangle $ 
can be easily proved to be strong potential functions for general constrained single-hop networks, provided that 
i) $P$ is a symmetric strictly positive definite matrix,
 ii)  $g(x)$ is $C^\infty[\errep\to \erre]$, increasing, 
null in the origin (i.e., $g(0)=0$), with null derivative in the origin (i.e., $g'(0)=0$), polynomial
asymptotic behavior for large $x$ (i.e., $\limsup_{x\to 
\infty} g^{(h_0)}<\infty$ for some $h_0 \in \enne$, and
 $\lim_{x\to \infty} \frac{g^{(h+1)}(x)}{g^{(h)}(x)}=0$ $\forall h<h_0$), 
 and such that $\lim_{x\to \infty} \frac{g(x)}{x}= \infty$.
Particularly relevant are functions in the form $ G(X)= \langle X^{\alpha+1}P\cdot X^{\alpha+1} \rangle $ with $\alpha>0$.
 

The following result allows us to more precisely characterize 
the class of well defined potential functions: 

\begin{corollario}
Given a weak (strong) non negative potential function $G_1(X)$ and
 a weak (strong) non negative monotonic  potential function 
  $G_2(X)$, then:
\begin{itemize}
\item $G(X)=\alpha G_1(X)+\beta G_2(X) \qquad \forall \alpha, \beta\ge0$
\item $G(X)=G_1(X) G_2(X)$
\label{algebra}
\end{itemize}
are weak (strong) potential functions.

Furthermore given $G(X)$, weak (strong)  potential function and 
$g(x)\in C^2[\errep \to \erre]$, ($g(x)\in C^\infty[\errep\to \erre]$), increasing
with at least linear and sub-exponential (polynomial) asymptotic behavior, i.e.
such that: 
$\liminf_{x\to \infty} \frac{g(x)}{x}>0$, 
$\lim_{x\to \infty}\frac{g(x+1)}{g(x)}=1$, 
$\lim_{x\to \infty}\frac{g'(x)}{g(x)}=0$, 
$\lim_{x\to\infty}\frac{g''(x)}{g'(x)}=0$ 
(or $\limsup_{x\to \infty}g^{(h_0)}(x)<\infty$, for some $h_0$ and 
$\lim_{x\to \infty}\frac{g^{(h+1)}(x)}{g^{(h)}(x)}=0$, $\forall h< h_0$),
then $g(G(X))$ is a weak (strong) potential function. 
If additionally $g'(0)=0$ also $G(g(X))$ is a weak (strong) potential function.
\end{corollario}

The proof, which consists in the verification that all conditions of the
statement of Theorem~\ref{max-scalar-theo-sh} (Theorem~\ref{max-scalar-theo-sh-strong}) are met,
is rather long and tedious even if conceptually 
straightforward. For these reasons, we omit it.

Previous corollary characterizes the algebraic structure of potentials and
makes the verification of throughput optimality easier for 
$\nabla G(X)$-{\em max scalar} policies associated with potentials with complex structure such as:
 $G(X) =\sum_m g(x^{(m)}) + \langle X^{1+\alpha} P \cdot X^{1+\alpha}\rangle $, 
 $G(X) =\sum_m g(x^{(m)}) \cdot \langle X^{1+\alpha}P \cdot X^{1+\alpha}\rangle $ or 
 $G(X)=g(\langle X^{1+\alpha}P\cdot X^{1+\alpha}\rangle )$,
 where $g(x)$ is a scalar potential and $P$ is a symmetric matrix with non negative entries.

The following Corollary allows us to further extend the class of throughput
optimal scheduling policies: 
\begin{corollario}

Given a weak (strong) potential function $G(X)$,
 any scheduling policy $\pi_{\imp}$ achieves the same throughput performance 
 (queue stability in non overloaded conditions) of the associated 
 $\pi_{\gmax}$ policy, if it satisfies the following property:
\begin{equation}
\lim_{\|X\|\to \infty} \langle  \nabla G(X)(I-R)^T \cdot (D^{\gmax}- D^{\imp})  \rangle 
= o(\|G(X)\|).  \label{corr-imp}
\end{equation}

\label{approx}
\end{corollario}
The proof is reported in Appendix \ref{app-proofs5}

In general, it is easy to see that scheduling policies according to which:
\begin{equation}
D_t^{\gmax}=\argmax_{\DSX}
\langle \nabla G(Z_t)(I-R)^T \cdot D\rangle 
\label{max-scalar-imp}
\end{equation}
meet constraint (\ref{corr-imp}) as long as $\EE[(\|Z_t-X_t\|)^h]$ is bounded for any $h\in \enne$.
Thus, the class of throughput optimal scheduling policies includes $\nabla
G(X)$-{\em max scalar} policies 
operating with imperfect/delayed queue status information as well as frame-based $\nabla
G(X)$-{\em max scalar} policies 
(i.e., policies in which the computation of a new departure vector is not executed at
every slot, but just once a while), etc.

\subsection{ Policies with memory}\label{sect-main-mem}

A further extension to the class of throughput optimal policies can be provided, 
considering scheduling policies with
memory~\cite{giaccone-shah,modiano-shah,linear-complexity-tassiulas}:
\begin{teorema}
Given a weak potential function for  the system of queues $G(X)$, 
satisfying:
\begin{equation}
\lim_{\|X\| \to \infty}
 \frac{\|H_G(X)\|^\beta} {\|\nabla G(X)\|}=0
\label{asymptder-memory}
\end{equation}
for some  $\beta>1$. The network of queues is $\|X\|^h$-stable for any $h\in \enne$
 under i.i.d. admissible arrival processes and static service constraints whenever
 a scheduling policy with memory $\pi_{\mem}$ is employed, provided that:
\begin{enumerate}
\item departure vectors selected by $\pi_{\mem}$ satisfy the following
monotonicity property:
\begin{equation}
\hspace{-4 mm}\langle \nabla G(X_{t+1})(I-R)^T \cdot D^{\mem}_{t+1} \rangle \; \; \ge \;\; \langle \nabla
G(X_{t+1})(I-R)^T \cdot D^{\mem}_{t}\rangle 
\label{propincr}
\end{equation}
at every $t$;
\item  for some $\delta>0$, the selected departure vector $D_t^{\mem}$ satisfies: 
$$D_{t}^{\mem}=\argmax_{\DX}\langle \nabla G(X_t)\cdot D(I-R)\rangle $$ 
 with a probability no smaller then $\delta$; in other words $D_{t}^{\mem}= D^{\gmax}_{t}$ with probability at least $\delta$, at every $t$. 
\end{enumerate}
\label{memory}
\end{teorema}
The proof is reported in Appendix \ref{app-proofs5}

As already mentioned  it is possible to simply  implement a  scheduling  policy satisfying  
 properties 1 and 2 in Theorem~\ref{memory}  by generating at random   an admissible candidate 
departure vector $D^c_t$, and  selecting  the departure vector $D^{\mem}_t$
according to the rule  $D^{\mem}_t=\argmax \{\langle X \cdot D^c_{t}\rangle , 
\langle X \cdot D^{\mem}_{t-1}\rangle \}$. 

\noindent{\bf Remark:} Observe that in this case the space state of the
DTMC representing the evolution of the system of queues must be properly
defined.  Information about the last employed 
departure vector must be, indeed, represented in the state.  A natural choice is to take $Y_t=[X_t, D^M_{t}]$ with $D^M_{t}=D^{\mem}_{t}  $. 
Further notice that (\ref{asymptder-memory}) is satisfied whenever $G(X)$ exhibits 
 a polynomial behavior for large $\|X\|$.

When $G(X)$ is a strong potential,
previous result can be extended under more general assumptions on arrival
processes and service constraints. In this
latter case however the complexity of the scheme significantly increases,
since the scheduling policy has to memorize the last selected departure
vector for every possible state $S\in \Sc^D$ of the  Markov
Chain representing service constraints evolution.

\begin{teorema}
Let $G(X)$ be a strong potential function of the system of queues. 
The network of queues is $\|X\|^h$-stable for any $h\in \enne$
 under admissible MMBP arrival processes and general service constraints,
whenever a scheduling policy with memory $\pi_{\mem}$ is employed, provided that:
\begin{enumerate}
\item at every time slot $t$, the following  property is satisfied by departure 
vectors selected by $\pi_{\mem}$:
\begin{equation}
\langle \nabla G(X_{t})(I-R)^T \cdot D^{\mem}_{t} \rangle \;\;\;
\ge \;\;\; \langle \nabla G(X_{t})(I-R)^T \cdot D_{t}^{M}(S^D_{t})\rangle ,
\label{propincrstrong}
\end{equation}
where $D^M_{t}(S^D_{t})$ is the departure vector employed by the scheduler $\pi_{\mem}$ at 
the last epoch $t_* < t$ in which $S^D_{t_*}=S^D_{t}$;

\item for some $\delta>0$
the selected departure vector $D^{\mem}_t$ satisfies: $$D^{\mem}_{t}=\argmax_{\DSX}\langle \nabla G(X_t)\cdot D(I-R)\rangle $$
with a probability no smaller then $\delta$.

\end{enumerate}

\label{memory-strong}
\end{teorema}
The proof is reported in Appendix \ref{app-proofs5}.
Observe that the  property expressed by (\ref{propincrstrong}) represents the natural 
extension of (\ref{propincr}) to the case dynamic constrains scenario.
To satisfy such property, the algorithm has to memorize
 the last selected departure vector $D^{\mem}_t(S^D)$, for every possible state $S\in \Sc^D$. Indeed 
(\ref{propincrstrong}) can be achieved by comparing, at time $t$ a randomly generated 
 candidate departure vector $D^c_{t}$  with the  memorized vector $D_{t}^{M}(S^D_{t})$.


\section{Conclusions}
\label{sect-concl}

The research on throughput optimal 
scheduling policies in constrained queuing networks has mainly focused on the
analysis of {\em max scalar} scheduling policies employing diagonal weights. Only
recently~\cite{meyn,bambos}, the existence of a class of throughput optimal {\em
max scalar} policies employing off-diagonal weights has been proved for arbitrary networks. In this paper, we have derived a general set of sufficient conditions 
for throughput optimality that lead to significant extension of  
results in~\cite{meyn,bambos}, defining a large body of non diagonal throughput optimal scheduling policies. Furthermore, we have shown, how low complexity scheduling policies with memory can achieve optimal throughput properties
 under general conditions (i.e., under non i.i.d. arrival processes and dynamic
 services constraints). This paper contributes to make a step toward full comprehension 
of the structure of throughput optimal scheduling policies in constrained queuing systems. 
The analysis of delay properties for  scheduling algorithms with off-diagonal weights is still an important challenging open issue.

\appendix  
\section{Proofs  of Theorems in Section \ref{sect-lyapu}}
\label{app-proofs3}

\profof{Theorem~\ref{th3}}

The fact that DTMC $Y_{t_k}$ is strongly stable, i.e., 
$
\limsup_{k\to \infty} \EE[\|X_{t_k}\|] <\infty
$
is an immediate consequence of (\ref{inside-sampled}) and (\ref{outside-sampled})~\cite{tweedie-meyn}.
Then, considering a generic instant $t$ and denoting by $T(t)= \max \{t_k\le t\}$, we have:
\[
 \EE[\|X_t\|]\le \EE[\|X_{T(t)}\|]+
\EE\left[\|\sum_{\tau =T(t)}^{t-1}\|A_\tau-D_\tau(I-R)\|\right]
\] 
where $\EE[\|\sum_{\tau =T(t)}^{t-1}A_\tau-D_\tau(I-R)\|]\le 
\EE[\sum_{\tau =T(t)}^{t-1}\|A_\tau-D_\tau(I-R)\|]\le \EE[t-T(t)]c$
where $c$ is an upper bound for $A_t-D_t(I-R)$ (which are bounded by assumption).
The assertion follows letting $t\to \infty$. Indeed
$\limsup_{t\to \infty}\EE[t-T(t)]<\infty$
as consequence of standard renewal arguments, since  $\{t_k\}$ is, by assumption, a sequence of non-defective  regeneration instants (i.e.
$\EE[(z_k)^2]=\EE[(t_{k+1}-t_k)^2]<\infty$). 

\vspace{0.5 cm}

\profof{Theorem~\ref{th4}}

Since the assumptions of Theorem~\ref{foster} are satisfied, every state of
the DTMC is positive recurrent and the DTMC is weakly stable.
In addition, to prove that the system is $F(X)$-stable, we shall show that
$
\lim_{t \rightarrow \infty} \sup \EE[F(X_t)]< \infty.
$

Let $\Hc_b$ be the set of values taken by $Y_t$, for which
$\|X_t\| \le b$ (where (\ref{outside-F}) does not apply).
It is immediate to see that $\Hc_b$ is a compact set.
Outside this compact set, Equation (\ref{outside-F}) holds, i.e.
\[
\EE[ \Lc(Y_{t+1}) - \Lc(Y_{t}) \mid Y_t] < -\epsilon F(X_t) \quad \forall Y_t\nin \Hc_b
\]
Averaging over  all $Y_t$'s that do not belong to $\Hc_b$, we obtain
\[
\EE[\Lc(Y_{t+1}) - \Lc(Y_{t}) \mid Y_{t} \nin \Hc_b] <
-\epsilon \EE[F(X_t) \mid Y_t \nin \Hc_b]
\]
Instead, for $Y_t\in \Hc_b$, since $\Hc_b$ is a compact set and $\Lc(Y)$
continuous we have:
\[
\sup_{Y_t\in \Hc_b} \EE[ \Lc(Y_{t+1}) \mid Y_{t} ] \le \max_{Y_t\in \Hc_b}
\Lc(Y_t)+v_0 < \infty. 
\]
Denoting by $c= \max_{Y_t\in \Hc_b} \Lc(Y_t)+v_0 $ and combining the two previous expressions, we obtain
\[
\begin{array}{c}
\vspace{1mm}
\EE[\Lc(Y_{t+1})] 
 < c \Prob\{Y_t\in \Hc_b\} +\Prob\{Y_t\nin \Hc_b\} \cdot \left\{
\EE[\Lc(Y_t) \mid Y_t \nin \Hc_b] -\epsilon \EE[F(X_t) \mid Y_t \nin \Hc_b]
\right\}
< \cr
\hspace{10mm}
\!\!\!< c + \EE[\Lc(Y_t)] - \epsilon \EE[F(X_t)] + c_0
\end{array}
\]
where $c_0$ is a constant such that
$c_0 \geq \{-\EE[\Lc(Y_t) \mid Y_t \in \Hc_b]+$ $ \epsilon \EE[F(X_t) \mid Y_t \in
\Hc_b]\}\Prob\{Y_t\in \Hc_b\}$.
Note that $c_0$ can be chosen finite, being $\Hc_b$ a compact set, 
and both $F(X)$ and
$\Lc(Y)$ continuous.

By summing over all $t$ from 0 to $\tau_0-1$, we obtain
\[
\EE[\Lc(Y_{\tau_0})] < \tau_0c +\EE[\Lc(Y_0)]-\epsilon \sum_{t=0}^{\tau_0-1}
\EE[F(X_t)]+\tau_0c_0
\]

Thus, for any $\tau_0$, we can write
\[
\frac{\epsilon }{\tau_0} \sum_{t=0}^{\tau_0-1} \EE[F(X_t)] < c + \frac{1}{\tau_0}
\EE[\Lc(Y_{0})] - \frac{1}{\tau_0}\EE[\Lc(Y_{\tau_0})]+c_0
\]

$\EE[\Lc(Y_{\tau_0})]$ is lower bounded by definition; assume $\EE[\Lc(Y_{\tau_0})] >
c_1$.
Hence
\[
\frac{\epsilon }{\tau_0} \sum_{t=0}^{\tau_0-1} \EE[F(X_t)] < c + \frac{1}{\tau_0}
\EE[\Lc(Y_0)] - \frac{c_1}{\tau_0} + c_0
\]
For $\tau_0\to \infty$, being $\EE[\Lc(Y_0)]$ and $c_1$ finite, we can write
\[
\limsup_{\tau_0 \rightarrow \infty} \frac{\epsilon }{\tau_0}
\sum_{t=0}^{\tau_0-1} \EE[F(X_t)] < c + c_0
\]
The assertion immediately follows.


\section{Proofs  of Theorems in Section \ref{sect-main}}
\label{app-proofs5}

Before proceeding with the proofs of the Theorems in Section~\ref{sect-main}, 
 we recall some standard consequences of Taylor Theorem, of which 
we will be make extensive use, and we prove three useful  Lemmas:

\begin{proposizione}
Let $G(X)$ $G: [\erre^{+M}\to \erre]$ be $h$-times continuously
differentiable over an open ball $\Bc$ centered at a vector $X$.
Then, for any $Y$ such that $X+Y \in \Bc$, 
\begin{equation}
 G(X + Y) = \sum_{i=0}^{h-1} \frac{1}{i!}Y^i(\partial^i G)(X) + R^{(h)}_G(X,Y) 
\label{taylor}
\end{equation}
where the $h$-order remainder $ R^{(h)}_G(X,Y)$ is given by:
$R^{(h)}_G(X,Y)= \frac{1}{h!} Y^h (\partial^h G)(X+\beta Y)$, for some $\beta=[0,1]$.
\label{taylor-P}
\end{proposizione}

In particular if $G(X)$ is twice continuously
differentiable over an open ball $\Bc$ centered at a vector $X$,
recalling that $\nabla G(X)$ denotes the gradient of G at $X$, and
$H_G(X)$ denotes the Hessian
of the function $G$ at $X$, 
for any $Y$ such that $X+Y \in \Bc$, we have:
\[
G(X + Y) = G(X) + R^{(1)}_G(X,Y) 
\]
with $R^{(1)}_G(X,Y) = \langle  \nabla G(X+ \beta Y) \cdot Y\rangle
$  for some $\beta \in [0, 1]$, and: 
\begin{equation}
 G(X + Y) = G(X) + \langle \nabla G(X) \cdot Y\rangle + R^{(2)}_G(X,Y)
\label{2orderexp}
\end{equation}
$R^{(2)}_G(X,Y)= \frac{1}{2} Y H_G(X+\beta Y) Y^T 
$ for some $\beta\in [0,1]$.
The above Taylor expansion can be generalized to vectorial functions applying (\ref{taylor}) component-wise.
In particular we will make use of the following result.
Given $G(X)$ twice continuously differentiable over an open ball $\Bc$ centered at a vector $X$, for any $Y$
such that $X+Y \in \Bc$,  and any $Z \in \erren$ we have:
\begin{equation}
\langle \nabla G(X+Y)\cdot Z \rangle = \langle \nabla G(X)  \cdot Z \rangle + R^{(1)}_{\nabla G}(X,Y,Z)
\label{gradientvar}
\end{equation}
with $R^{(1)}_{\nabla G}(X,Y,Z)= \langle ( \nabla  \langle \nabla G(X+\beta Y)\cdot Z \rangle) \cdot Y\rangle = 
  \frac{1}{2} Z H_G(X+\beta Y) Y^T 
$ for some $\beta \in [0, 1]$.


\begin{lemma}
If $G(X)$ satisfies conditions of Theorem \ref{max-scalar-theo-sh}, then:
\[
\lim_{\|X\| \to \infty}  \langle \nabla G(X) \cdot  \tilde{X} \rangle =\infty
\]
$\tilde{X}$ being the normalized vector parallel to $X$
\label{lemma2-a}
\end{lemma}

\proof
The proof can be immediately obtained by applying l'Hopital's rule to the indefinite
form~(\ref{asympG}):
\[
\lim_{\alpha \to\infty } \frac{G(X)}{\|X\|}=
\lim_{\alpha \to\infty } \frac{G( \alpha \tilde{X})}{\alpha}
\]
and recalling that $\lim_{\alpha \to\infty }  
\langle \nabla G(\alpha \tilde{X}) \cdot  \tilde{X} \rangle
=\lim_{\| X \| \to\infty }  \langle \nabla G(X) \cdot  \tilde{X} \rangle$
 exists in light of  (\ref{asymptderadd}). 
Observe  as immediate consequence  of previous  statement we get:
\[ 
\lim_{\| X \| \to\infty } \| \nabla G(X) \|=\infty
\]


\begin{lemma}
If $G(X)$ satisfies the conditions of Theorem \ref{max-scalar-theo-sh} 
 then:
\begin{equation}
G(X+Y)-G(X)=R^{(1)}(X,Y) =
\left\{ \begin{array}{l} O(\|\nabla G(X)\|)\\
                         o(G(X)) \end{array}   \right.      
 \qquad \mbox{as } \|X\|\to \infty, 
\label{nuova1}
\end{equation}
whenever $Y$ is an arbitrary bounded  vector. 
If  $G(X)$ satisfies the conditions of Theorem \ref{max-scalar-theo-sh-strong}, then: 
\begin{equation}
\EE[G(X+Y)]- G(X)= \EE[R^{(1)}_G(X,Y)] =\left\{ \begin{array}{l} O(\|\nabla G(X)\|)\\
                         o(G(X)) \end{array}   \right.   
\qquad \mbox{as } \|X\|\to \infty , 
\label{nuova1a}
\end{equation}
whenever $Y$ is a random vector with finite polynomial 
moments  $\EE[\|Y\|^h]<\infty$ $\forall h$.

Similarly, 
if $G(X)$ satisfies the conditions of Theorem~\ref{max-scalar-theo-sh}, then:
\begin{equation}
 \begin{split}
\langle \nabla G(X+Y),Z \rangle - \langle \nabla G(X), Z \rangle =  R^{(1)}_{\nabla G}(X,Y,Z)=\left\{ \begin{array}{l} O(\|H_G(X)\|)\\
                         o(\| \nabla G(X)\|) \end{array}   \right.     
\qquad \mbox{as } \|X\|\to \infty ,  
\end{split}
\label{nuova2}
\end{equation}
whenever $Z$ and $Y$ are two arbitrary bounded vectors.
If $G(X)$ satisfies the conditions of Theorem~\ref{max-scalar-theo-sh-strong}, then:
\begin{equation}
\begin{split}
\EE\langle \nabla G(X+Y),Z \rangle - \langle \nabla G(X), Z \rangle = \EE[R^{(1)}_{\nabla G}(X,Y,Z)]=\left\{ \begin{array}{l} O(\|H_G(X)\|)\\
                         o(\| \nabla G(X)\|) \end{array}   \right.                                                                         
\qquad \mbox{as } \|X\|\to \infty ,  
\end{split}
\label{nuova2a}
\end{equation}
whenever $Z$ is an arbitrary bounded vector and  
$Y$ is a   random vector with finite polynomial moments 
(i.e., $\EE[\|Y\|^h]<\infty$, $\forall h)$.

At last, if $G(X)$ satisfies the conditions of Theorem~\ref{max-scalar-theo-sh}, then:
\begin{equation}
R^{(2)}(X,Y)= O( \|  H_G(X)\| )  \qquad R^{(2)}(X,Y) =o( \|  \nabla G(X)\| )
\label{nuova3}
\end{equation}
for any  vector $Y$. If  $G(X)$ satisfies the conditions of Theorem~\ref{max-scalar-theo-sh-strong}, then:
 \begin{equation}
\EE[R^{(2)}(X,Y)]=  O( \|  H_G(X)\| ) \qquad \EE[R^{(2)}(X,Y)]=o( \|  \nabla G(X)\| )
\label{nuova3a}
\end{equation}
whenever $Y$ is a random vector with finite polynomial 
moments (i.e., $\EE[\|Y\|^h]<\infty$, $\forall h$).

\label{lemma3}
\end{lemma}

\proof

Properties (\ref{nuova1}) and (\ref{nuova2})  are an immediate consequence of the sub-exponential
 behavior  of $G(X)$, i.e (\ref{asymptderadd}).
Now focusing on  (\ref{nuova1a}), observe that 
expanding $G(X)$ in Taylor series around $X$, we obtain:
\[
 \EE[G(X + Y)] = G(X) +
 \EE[\sum_{i=1}^{h_0-1} \frac{1}{i!}Y^i(\partial^i G)(X)] + \EE[R^{(h_0)}_G(X,Y)] 
\]
where $\EE[R^{(h_0)}_G(X,Y)]=\frac{1}{h_0!} \EE[Y^{h_0} (\partial^{h_0} G)(X+\beta Y)]
 \le \frac{1}{h_0!} \EE[\|Y^{h_0}\|] \sup_{Z\in \erre^{+M}} \|(\partial^{h_0} G)(Z)\|< \infty $, 
because  by assumptions  $\EE[Y^{h_0}]$ is bounded as well as 
 $\sup_{Z\in \erre^{+M}} \|(\partial^{h_0} G)(Z)\|< \infty $ (recalling (\ref{asymptder-strong2})).  
Thus the last term is negligible with respect to $G(X)$ and    $\| \nabla G(X)\|$ since 
both  $G(X) \to \infty$  (by hypothesis) and   $\|\nabla G(X)\| \to \infty$  (by  Lemma $\ref{lemma2-a}$)  as 
$\|X\| \to \infty$ b).  Furthermore,  $\EE[\sum_{i=1}^{h_0-1} \frac{1}{i}Y^i(\partial^i G)(X)]=
\sum_{i=1}^{h_0-1} \frac{1}{i!} \EE[Y^i](\partial^i G)(X)= O(\|\nabla G(X) \|)= o( G(X))$,
since $\EE[Y^i]<\infty$ and $\|(\partial^i G)(X)\|= o(\partial^{i-1} G(X))$ for any $1\le i < h_0$,
  from  (\ref{asymptder-strong}). Thus (\ref{nuova1a}) is proved. 
(\ref{nuova2a}) can be proved  repeating the same arguments to every 
component of  $\nabla G(X)$.    

(\ref{nuova3})  can be proved  observing that  by  definition   $R^{(1)}_{\nabla G} (X,Y,Y)$ and  $R^{(2)}_{ G} (X,Y)$ are closely related, indeed:
  $R^{(1)}_{\nabla G} (X,Y,Y)= Y H_G(X+\beta_1 Y)  Y^T$ for a $\beta_1 \in [0,1]$, while 
 $R^{(2)}_{\nabla G} (X,Y)= Y H_G(X+\beta_2 Y)  Y^T$ for a $\beta_2 \in [0,1]$,  possibly different from $\beta_1$. 
Now by   (\ref{asymptderadd})  we get that  $\lim_{\|X \| \to \infty}   \frac{R^{(2)}_{G} (X,Y)  } {R^{(1)}_{\nabla G} (X,Y,Y)}= \lim_{\|X \| \to \infty}   \frac{Y H_G(X+\beta_2 Y)  Y^T } {Y H_G(X+\beta_1 Y)  Y^T}=1$,  furthermore  from
(\ref{nuova2}) we have
$\lim_{\|X \| \to \infty} \frac{ R^{(1)}_{\nabla G} (X,Y,Y)}{\| \nabla G(X)\| }=0  $,  (or in alternative $\liminf_{\|X \| \to \infty} \frac{ R^{(1)}_{\nabla G} (X,Y,Y)}{\| H_G(X)\| }>0  
$ and $\limsup_{\|X \| \to \infty} \frac{ R^{(1)}_{\nabla G} (X,Y,Y)}{\| H_G(X)\| }<\infty $);
 thus  combining both we get: 
  $\lim_{\|X \| \to \infty}   \frac{R^{(2)}_{G} (X,Y)  } {\| \nabla G(X)\| }=0$ (or in alternative
$\liminf_{\|X \| \to \infty} \frac{R^{(2)}_{\nabla G}  (X,Y)}{\| H_G(X)\| }>0  $ and $\limsup_{\|X \| \to \infty} \frac{R^{(2)}_{ G} (X,Y)}{\| H_G(X)\| }<\infty $)

At last  (\ref{nuova3a})  can be proved observing that:
$\EE[G(X + Y)] =   G(X) + \langle \nabla G(X) \cdot \EE[Y] \rangle +\EE[R^{(2)}_G(X,Y)] = 
G(X) + \langle \nabla G(X) \cdot \EE[Y] \rangle + \EE[\sum_{i=2}^{h_0-1} \frac{1}{i!}Y^i(\partial^i G)(X)] + \EE[R^{(h_0)}_G(X,Y)]$ thus:  
\[
\EE[R^{(2)}_G(X,Y)] = \EE[\sum_{i=2}^{h_0-1} \frac{1}{i!}Y^i(\partial^i G)(X)] + \EE[R^{(h_0)}_G(X,Y)]
\]
Now  from  (\ref{asymptder-strong2}) and  (\ref{asymptder-strong}), as before,  we can conclude that all terms on the right are  $O(\| H_G G(X) \|)$ and  $o(\| \nabla G(X) \|)$.
\endproof

\begin{lemma}
If $G(X)$ satisfies the conditions of Theorem \ref{max-scalar-theo-sh} (and in
particular condition (\ref{negder})), then: 
\begin{equation}
 \max_{\DX}
\langle \nabla G(X_t)(I-R)^T \cdot D \rangle \;\;\ge \; 
\max_{ D \in \Dc } \langle \nabla G(X_t)(I-R)^T \cdot D \rangle + o( \nabla G(X_t) )
\label{max-scalar-prop}
\end{equation}
\label{lemma2}
\end{lemma}
i.e, there is always an ``almost'' optimal feasible departure vector satisfying the 
condition $D_t\le X_t$ among the departure vectors that maximize the scalar
product $\langle \nabla G(X_t)(I-R)^T \cdot D\rangle $. 

\proof
We denote by $\tilde{D}= \argmax_{ D \in \Dc } \langle \nabla 
G(X_t)(I-R)^T \cdot D \rangle $, and by 
$D^{*}= \min(\tilde{D}, X_t )$. 
Observe that $\tilde{D}$ can be always assumed to be feasible, since by assumption every vertex  of $\Dc$ corresponds  by assumption  to a feasible vector.
As a consequence also  $D^{*}$ is, by   construction,  feasible.
Note that 
\begin{equation}
\langle \nabla G (X_t)(I-R)^T \cdot D^{*}
\rangle \; \;\le\;\; \langle \nabla G (X_t)(I-R)^T \cdot D_t^{\gmax} \rangle 
\label{prooflemma_eq1}
\end{equation}
since by construction $D^*$ is feasible,  $D^{*} \in \Dc$ and   $D^{*} \le X_t$.

Furthermore note that by construction $X^{*}_t =X_t- D^{*}$ and 
$\tilde{D}- D^{*}$ are orthogonal since the non null components 
of $\tilde{D}- D^*=\max(\tilde{D}-X_t, 0)$
coincide with the null of $X^{*}_t= \max(X_t- \tilde{D}, 0)$. Thus according to
(\ref{negder}): 
\begin{equation}
\langle \nabla G (X^{*}_t)\cdot (\tilde{D}- D^*)(I-R)\rangle =
\langle \nabla G (X^{*}_t)(I-R)^T\cdot \tilde{D}- D^*\rangle 
\;\le 0
\label{prooflemma_eq2}
\end{equation}
 now expanding in Taylor series $ \nabla G (X)$ around point $X_t$ we obtain
\[
\nabla G(X^{*}_t)= \nabla G(X_{t}) + R^{(1)}_{\nabla G}(X_t,-D^{*})
\]
Since  $D^{*}$ is bounded in norm, from~(\ref{nuova2})  we can
 conclude that the remainder $R^{(1)}_{\nabla G}(X_t,-D^{*}_t)$ is $o(\nabla G (X_t))$ and thus:
 \begin{equation}
 \langle \nabla G (X_t) \cdot (\tilde{D}- D^{*})(I-R)\rangle =
 \langle \nabla G (X^{*}) \cdot (\tilde{D}- D^{*})(I-R) \rangle + o(\nabla G (X_t))
 \label{prooflemma_eq3}
 \end{equation}
 from which the assertion follows recalling (\ref{prooflemma_eq1}) and (\ref{prooflemma_eq2}). Indeed
 \begin{multline*}
 \langle \nabla G (X_t) \cdot (\tilde{D}- D_t^{\gmax})(I-R) \rangle \;\; 
\stackrel{(\ref{prooflemma_eq1})}{\le} \langle \nabla G (X_t) \cdot (\tilde{D}- D^{*})(I-R)
  \rangle\\ \stackrel{(\ref{prooflemma_eq3})}{=}\
 \langle \nabla G (X^{*}_t) \cdot (\tilde{D}- D^{*})(I-R)\rangle + o(\nabla G (X_t))
 \stackrel{(\ref{prooflemma_eq2})}{\le}  o(\nabla G (X_t))
\end{multline*} 


\vspace{0.5 cm}

\profof{Theorem~\ref{max-scalar-theo-sh}}

First, observe that since arrivals are assumed i.i.d. and service constraints are
assumed to be static, we have $\Hc=\Xc$.

The idea of the proof is rather simple; $G(X)$ can be interpreted as a
 Lyapunov function for the system.
The stability of the network of queues follows from the fact that
drift conditions of Theorem \ref{th4} are verified.

First, we evaluate the drift of $G(X_t)$ for large values of $X_t$. By
definition:
\[
\Delta \Lc =\EE[G(X_{t+1})- G(X_t)\mid X_t]=\EE \left[G\left(X_{t}+A_t-D_t^{\gmax}(I-R)\right) \big| X_t \right]- G(X_t)
\]
and approximating $G(X_t+A_t-D_t^{\gmax}(I-R))$ with its first order Taylor polynomial centered at
$X_t$, we get:
\begin{multline}
 \left[G\left(X_{t}+A_t-D_t^{\gmax}(I-R)\right) \big| X_t \right]\\
 =G(X_{t})+ \langle \nabla G(X_t)\cdot \left[\left(A_t-D_t^{\gmax}(I-R)\right)\right]\rangle + R^{(2)}_{G}\left(X_t, A_t-D_t^{\gmax}(I-R)\right)
\end{multline}
where  the remainder 
 $R^{(2)}_{G}(X_t, A_t-D_t^{\gmax}(I-R))$,  satisfies:
\[
\lim_{\|X_t\|\to \infty} \frac{\Big|\Big|R^{(2)}_{G}\left(X_t,A_t-D_t^{\gmax}(I-R)\right)\Big|\Big|}{\|\nabla G(X_t)\|}=0
\]
 in light of $(\ref{nuova3})$  (Lemma~\ref{lemma3}), since  both $A_t$ and $D_t^{\gmax}$ are  bounded norm   vectors.
Thus:
\begin{multline}
\EE\left[G\left(X_{t}+A_t-D_t^{\gmax}(I-R)\right)\big| X_t \right]=\\
 G(X_{t})+ \langle \nabla G(X_t)\cdot \EE\left[\left(A_t-D_t^{\gmax}(I-R)\right)\right]\rangle + o(\|\nabla G(X_t)\|)
\label{appproxGdrift}
\end{multline}
with, 
$$ \langle \nabla G(X_t)\cdot \EE[(A_t-D_t^{\gmax}(I-R))]\rangle = \langle \nabla G(X_t)\cdot
\Lambda-D_t^{\gmax}(I-R) \rangle =\langle \nabla G(X_t)\cdot
\Lambda\rangle - \langle \nabla G(X_t)\cdot D_t^{\gmax}(I-R)\rangle.$$ 
Since by assumption $\Lambda(I-R)^{-1}$ lies in
the interior of $\Dc$, an
$\epsilon'>0$ can be found, such that also $\Lambda(I-R)^{-1}+\epsilon'
\tilde{D}$  lies in $\Dc$,   with $\tilde{D}=\argmax_{\Dc} \langle \nabla G(X_t)\cdot D(I-R)\rangle$. 

we obtain:
\begin{multline}
\langle \nabla G(X_t)\cdot D_t^{\gmax}(I-R)\rangle = \max_{\DX}
\langle \nabla G(X_t)(I-R)^{T} \cdot D\rangle = \max_{\Dc} 
 \langle \nabla G(X_t)(I-R)^T \cdot D \rangle + 
o(\|\nabla G(X_t)\|) 
\label{drift1}
\end{multline}
where the second equation holds by virtue of Lemma~\ref{lemma2};
now:
\begin{multline}
 \max_{\Dc} \langle \nabla G(X_t)(I-R)^T \cdot D \rangle  \ge 
 \; \; \langle \nabla G(X_t)(I-R)^T \cdot \Lambda(I-R)^{-1}+ \epsilon'
\tilde{D} \rangle \ge
 \langle \nabla G(X_t) \cdot \Lambda\rangle + \epsilon\|\nabla G(X_t)\| 
\label{drif2}
\end{multline}
for a suitable $\epsilon>0$.  In particular  last equality  is  consequence of   \eqref{posorien}  and the definition of $\tilde{D}$, in light of which, we can claim
$\frac{\langle \nabla G(X_t)(I-R)^T \cdot \tilde{D} \rangle}{\|\nabla G(X_t) \|}=
\frac{\epsilon}{\epsilon'} $  for some $\epsilon>0$.
Now, combining \eqref{drift1} and \eqref{drif2}  we obtain:
\begin{multline}
 \langle \nabla G(X_t)\cdot D_t^{\gmax}(I-R)\rangle \ge  \langle \nabla G(X_t) \cdot \Lambda\rangle + \epsilon\|\nabla G(X_t)\| + o(\|\nabla G(X_t)\|) 
\label{drift}
\end{multline}

In conclusion:
\[
E\left[G(X_{t+1})- G(X_t)\mid X_t \right] \le - \epsilon \|\nabla G(X_t)\| + o(\|\nabla G(X_t)\|)
\] 
for large $X_t$, and therefore \eqref{outside-F} is satisfied, since
for any $\epsilon''<\epsilon$,  a sufficiently large $b>0$ can be found such that: 
\[
E\left[G(X_{t+1})- G(X_t)\mid X_t \right] \le - \epsilon'' \|\nabla G(X_t)\| 
\label{final-drift}
\]
for $\|X_t\|> b$.

{ 
Furthermore for any $X_t:\|X_t\|\le b$, 
$G(X_{t+1})-G(X_t)= G(X_{t}+A_t-D_t^{\gmax}(I-R))-G(X_t)$ is bounded. Indeed  once again we recall vector  $\| A_t-D_t^{\gmax}(I-R)\| $ is bounded in norm.
 Let    $\zeta$ be  a bound for $\| A_t-D_t^{\gmax}(I-R)\|$. 
Now  $\|X_{t+1}\|=\|  X_t + A_t-D_t^{\gmax}(I-R) \|  \le  
\|  X_t \| + \| A_t-D_t^{\gmax}(I-R) \| \le b+ \zeta$.

Thus being  $G(X)$ continuous,  and thus bounded over compact domains both from above and below:
$ G(X_{t+1})-G(X_t)\le \max_{X_t: \| X_t\|\le b+\zeta}G(X) - \min_{X_t: \| X_t\|\le b}G(X)$.
}
The $\|\nabla G(X)\|$-stability of the system of queues immediately follows since 
 $\lim_{\|X\|\to \infty} \nabla G(X)=\infty$ (as result of Lemma~\ref{lemma2-a})


\vspace{0.5 cm}

\profof{Theorem \ref{max-scalar-theo-sh-strong}}

The generalization to the case in which $S_t$ is a non trivial Markov Chain
can be carried out by sampling the process $Y_t$ in correspondence of the instants $\{t_k \}$
at which $S_{t_k}=S_0$ for some specific state $S_0$.
From theory of DTMC (recalling that $S_t$ has a finite number of states) immediately follows that 
 $\{t_k\}$ forms a sequence of non-defective
regeneration times for the system. 
Thus applying Corollary~\ref{th3a} we can prove the stability of the system of
queues. 
 To simplify the notation we assume traffic to be single-hop along our proof; 
 however we wish to emphasize that the proof for the more general case goes
 exactly along the same lines and can easily recovered by replacing the departure vector at time $t$, $D_t^{\gmax}$ with $D_t^{\gmax}(I-R)$
 in the following derivation.

Again we select $G(X)$ as Lyapunov function. Approximating $G(X)$ 
with its  second order Taylor expansion,
we get:


\begin{multline}
\EE [G(X_{t_{k+1}})\mid Y_{t_k}] \\= 
G(X_{t_{k}})+ < \nabla G(X_{t_k})\cdot \EE\left[ \sum_{t=t_{k}}^{t_{k+1}-1}
\big( A_t-D_t^{\gmax}\Big)\right]>+
\EE \left[ R^{2}_G\left(X_{t_k}, \sum_{t=t_{k}}^{t_{k+1}-1}(  A_t-D_t^{\gmax})\right)\right] 
\end{multline}
Now, since  all polynomial moments  of  vector 
$\sum_{t_{k}}^{t_{k+1}-1}(  A_t-D_t^{\gmax})$ are, by construction,  finite,  
(this because   
every vector   $ A_t-D_t^{\gmax} $ is bounded  in norm and polynomial moments of $z_k=t_{k+1}-t_{k}$ are finite,)  
 from (\ref{nuova3a}) we obtain that  $\EE\left[ R^{2}_G\left(X_{t_k},  \sum_{t_{k}}^{t_{k+1}-1}(  A_t-D_t^{\gmax})\right)\right]= o(\| \nabla G(X_{t_k})\|)$, i.e.,  

\begin{multline}
\EE [ G(X_{t_{k+1}})\mid Y_{t_k}]
=G(X_{t_{k}}) + \langle \nabla G(X_{t_k})\cdot \EE\left[\sum_{t=t_{k}}^{t_{k+1}-1}\Big(
A_t-D_t^{\gmax}\Big)\right]\rangle 
 +o(\|\nabla G(X_{t_k})\| )
\label{eq-strong-G}
\end{multline}

 
Furthermore:
 \begin{multline}
\langle \nabla G(X_{t_k})\cdot \EE\left[\sum_{t=t_{k}}^{t_{k+1}-1}\Big(
A_t-D_t^{\gmax}\Big)\right]\rangle =
\langle \nabla G(X_{t_k})\cdot  \EE\left[\sum_{t=t_{k}}^{t_{k+1}-1}\Big( A_t-D_{t_k}^{\gmax}+ D_{t_k}^{\gmax}-
D_t^{\gmax}\Big)\right]\rangle \\= \langle \nabla G(X_{t_k})\cdot \EE\left[\sum_{t=t_{k}}^{t_{k+1}-1}\Big(
A_t-D_{t_k}^{\gmax}\Big)\right]\rangle 
+
 \langle \nabla G(X_{t_k})\cdot \EE\left[\sum_{t=t_{k}}^{t_{k+1}-1}\Big( D_{t_k}^{\gmax}- D_t^{\gmax}\Big)\right]\rangle 
\label{driftstrong-1}
 \end{multline}
with: 
\begin{multline}
\langle \nabla G(X_{t_k})\cdot \EE\left[\sum_{t_{k}}^{t_{k+1}-1}( A_t-D_{t_k}^{\gmax})\right]\rangle = 
 \langle \nabla G(X_{t_k})\cdot \EE[z_k](\Lambda-D_{t_k}^{\gmax})\rangle \; \; \\ \le -
 \epsilon \EE[z_k] \|\nabla G(X_{t_k})\|
\label{driftstrong}
 \end{multline}
where the (first) equality follows from  classical reward-renewal arguments,   while
the following inequality is obtained with similar  arguments as in proof of Theorem \ref{max-scalar-theo-sh}. In particular  observe that 
 $  \langle \nabla G(X_t)\cdot \Lambda-D_t^{\gmax}\rangle =
\langle \nabla G(X_t)\cdot \Lambda\rangle - \langle \nabla G(X_t)\cdot D_t^{\gmax}\rangle$,
 and  since by assumption $\Lambda$ lies in the interior of $\Dc$, an
$\epsilon'>0$ can be found, such that also $\Lambda+\epsilon'
\tilde{D}$  lies in $\Dc$, with $\tilde{D}= \argmax_{\Dc} \langle \nabla G(X_{t_k})\cdot D\rangle$.
 Now, recalling Lemma \ref{lemma2}  we have:
$ \langle \nabla G(X_t)\cdot D_t^{\gmax}\rangle = \max_{\DX}
\langle \nabla G(X_t) \cdot D\rangle =
\max_{ D \in \Dc } \langle \nabla G(X_t) \cdot D \rangle + 
o(\|\nabla G(X_t)\|)\;\;  
\ge \; \; \langle \nabla G(X_t) \cdot \Lambda+ \epsilon'
\tilde{D} \rangle + o(\|\nabla G(X_t)\|) \ge \langle \nabla G(X_t) \cdot \Lambda\rangle + \epsilon \|\nabla G(X_t)\|$,
 where last inequality follows from (\ref{posorien}).

 \begin{multline}
\langle \nabla G(X_{t_k})\cdot \EE\left[\sum_{t=t_{k}}^{t_{k+1}-1}
\Big( D_{t_k}^{\gmax}- D_t^{\gmax}\Big)\right]\rangle =
\EE\left[\sum_{t=t_{k}}^{t_{k+1}-1}\langle \nabla G(X_{t_k}) \cdot D_{t_k}^{\gmax}- D_t^{\gmax}
\rangle \right]\\= 
\EE\left[ \sum_{t=t_{k}}^{t_{k+1}-1}\langle \nabla G(X_{t_k})- \nabla G(X_{t}) + \nabla G(X_{t}) \cdot  D_{t_k}^{\gmax}- D_t^{\gmax} \rangle \displaystyle \right] \\
= \EE\left[\sum_{t=t_{k}}^{t_{k+1}-1} \langle \nabla G(X_{t_k})- \nabla G(X_{t}) \cdot  D_{t_k}^{\gmax}- D_t^{\gmax}\rangle \displaystyle \right]+
 \EE\left[\sum_{t=t_{k}}^{t_{k+1}-1}\langle \nabla G(X_{t}) \cdot D_{t_k}^{\gmax}- D_t^{\gmax} \rangle \right]
 \end{multline}
Now $ \langle \nabla G(X_{t_k})-  \nabla G(X_{t}) \cdot D_{t_k}^{\gmax}- D_t^{\gmax}\rangle =o( \|\nabla
G(X_{t_k})\|) $  as  an immediate  consequence of  (\ref{nuova2}); in this regard, we recall that by hypothesis polynomial moments
 $\EE[\|X_{t_k}-X_{t}\|^h]$  are finite for any $h$; this again because  $\{t_k\}$ is a non-defective sequence of stopping times  and  arrival vector is bounded. 
At last observe that  the term $\langle \nabla G(X_{t}) \cdot D_{t_k}^{\gmax}- D_t^{\gmax} \rangle= \langle  \nabla G(X_{t})\cdot D_{t_k}^{\gmax}   \rangle-
 \max_{\Dc}   \langle  \nabla G(X_{t})\cdot D   \rangle + o(\| \nabla G(X_{t})\|)$  in light of Lemma \ref{lemma2}  with  $\langle  \nabla G(X_{t})\cdot D_{t_k}^{\gmax}   \rangle-\max_{\Dc}   \langle  \nabla G(X_{t})\cdot D   \rangle \le 0$

As a conclusion, recalling \eqref{nuova2a},  we have:
\[
\EE[G(X_{t_{k+1}})\mid X_{t_k}]-G(X_{t_{k}})\le - \epsilon \EE[z_k]\ \|\nabla G(X_{t_k})\| + o( \|\nabla G(X_{t_k})|)
\]
Therefore (\ref{outside-sampledF}). is satisfied, since for any 
$\epsilon''<\epsilon \EE[z_k]$, a $b>0$ can be found such that 
\[
E\left[G(X_{t_{k+1}})- G(X_{t_k})\mid X_t \right] \le - \epsilon'' \|\nabla G(X_{t_k})\| 
\]
for $\|X_t\|> b$.

At last, to show that (\ref{inside-sampledF}) is satisfied too, observe that for any $Y_{t_k}: \| X_{t_k} \| \le b$: 
\begin{multline*}
\EE\Big[G(X_{t_{k+1}})\Big]= \EE\left[G\Big(X_{t_k}+\sum_{t=t_{k}}^{t_{k+1}-1}(A_t-D_t^{\gmax})\Big)\right] \\
\stackrel{(\ref{eq-strong-G})}{=}
G(X_{t_{k}})+ \langle \nabla G(X_{t_k})\cdot \EE\left[\sum_{t=t_{k}}^{t_{k+1}-1}
\Big( A_t-D_t^{\gmax}\Big) \right]\rangle 
 + \sum_{i=2}^{h_0-1} \frac{1}{i!} \EE\left[ \left(\sum_{t=t_{k}}^{t_{k+1}-1}
 \Big(A_t-D_t^{\gmax}\Big)\right)^i \right]
 (\partial^i G)(X_{t_k}) \\+
 \frac{1}{h_0!} \EE\left[ \left(\sum_{t=t_{k}}^{t=t_{k+1}-1} \Big(A_t-D_t^{\gmax}\Big)\right)^{h_0}
  (\partial^{h_0} G)\left(X_{t_k}+\ \alpha \sum_{t=t_{k}}^{t_{k+1}-1} \Big(A_t-D_t^{\gmax}\Big)\right) \right]
 \end{multline*}
 can be easily shown to be bounded by $G(X_{t_k})+v_0$ for an appropriate $v_0>0$, 
since i) $ \EE\left[ \left(\sum_{t_{k}}^{t_{k+1}-1}
(A_t-D_t^{\gmax})\right)^i\right]$ are bounded for every $i$, as before; 
 ii) $G(X_{t_k})$ and its derivatives $(\partial^i G)(X_k)$ are by
 assumption bounded over compact domains   (in particular they are bounded over the domain $X: \|X\|\le b$) 
 because $G(X)\in C^\infty[\erre^M\to
 \erre]$;
 iii) $ (\partial^{h_0} G)(X_{t_k}+\ \alpha \sum_{t_{k}}^{t_{k+1}-1} (A_t-D_t^{\gmax}) )$ is
 bounded as before in light of (\ref{asymptder-strong2}).
The $\| \nabla G(X)\|$-stability of the system of queues immediately follows from Corollary~\ref{th3a} since $\lim_{\|X\|\to \infty} \nabla G(X)=\infty$ 
(as result of Lemma~\ref{lemma2-a}).

\vspace{0.5 cm}

\profof{Corollary \ref{strong-stab}}

Consider the Lyapunov function $\Lc(X)= \frac{1}{h+1}G(X)^{h+1}$; 
denoting by  $Z_t=A_t-D_t(I-R))$:
\[
G(X_{t+1})= G(X_{t}+Z_t)= G(X_t)+ \langle \nabla G(X_t) \cdot Z_t \rangle + R^{2}(X_t, Z_t)
\] 
Now recalling (\ref{nuova1}) and  (\ref{nuova3}), since by construction $Z_t$ is a bounded in norm vector
  we can claim that:   $\langle \nabla G(X_t) \cdot Z_t \rangle=o(G(X_t)) $,
and  $R^{2}(X_t, Z_t)=o(\| \nabla G(X_t))\|)$ as $X_t \to \infty$
Thus:

\begin{multline*}
\EE\left[\Lc(X_{t+1})\mid X_t\right]= \frac{1}{h+1} \EE \big[\big(G(X_t+Z_t)\big)^{h+1}\big]\\ =
 \frac{1}{h+1} \EE \left[ \Big( G(X_t)+ \langle\nabla G(X_t) \cdot Z_t \rangle + o(\| \nabla  G(X) \|) \Big)^{h+1}  \right]\\
= \frac{1}{h+1}  \left[ \big(G(X_t)\big)^{h+1}+ (h+1)\langle \nabla G(X_t) \cdot \EE [Z_t ]\rangle   \big( G(X_t)\big)^h  + o( \| \nabla  G(X) \|  \big(G(X_t)\big)^h \right]
\end{multline*}

Now considering $X_t$ sufficiently large, such that $G(X_t)$ is positive 
 (we recall that $G(X)\to \infty$, for $\|X\|\to \infty$ and thus, it must be
 positive outside some compact set),  from \eqref{drift},  we have:
\[
 (G(X_t))^h \langle \nabla G(X_t)\cdot \Lambda_t-D_t^{\gmax}(I-R)\rangle \;\; \le  -\epsilon (G(X_t))^h\|\nabla G(X_t) \|
 \]
for some $\epsilon>0$.

The $\|X\|^h$-stability  immediately follows,  observing that i) by construction $\lim_{ \|X\|\to \infty
} \frac{(G(X))^h \|\nabla G(X) \|}{\|X^h\|}=\infty$;  ii)  for any $X_t:\|X_t\|\le b$, 
 $\frac{1}{h+1}\Big[G\big(X_t+Y_t)\big)^{h+1}-\big(G(X_t)\big)^{h+1}\Big]$
can be bounded by an appropriate constant $v_0$ (this because $Y_t$ is bounded as well as $G()$ is bounded (from above and below) over compact sets); 

The extension  to the more general case  can be carried out  by observing that   $\Lc(X)=\frac{1}{h+1}G(X)^{h+1}$ is a  strong potential provided that   of  $G(X)$  is a strong  potential by Corollary \ref{algebra}.
Indeed denoting with  $Z_{t_k}= \sum_{t_k}^{t_{k+1}-1} \big(A_t-D_t(I-R)\big)$, we have:
\[
\EE \left[\Lc(X_{t_{k+1}})\mid X_{t_k}\right] =\Lc(X_{t_k}) + \EE [\langle \nabla  \Lc(X_{t_k}) \cdot Z_{t_k} \rangle] + \EE [R^{2}_{\Lc}(X_{t_k},Z_{t_k})] 
\]
with $\EE [R^{2}_{\Lc}(X_{t_k},Z_{t_k})]= o(\nabla \Lc(X_{t_k}))) $ 
 from 
(\ref{nuova3a}), 
since, by construction, all polynomial moments of $Z_{t_k}$ are finite. 

Now  observing that $\nabla  \Lc(X_{t_k})= \big(G(X_{t_k})\big)^h  \nabla G(X_{t_k})$,  we get: 
\[
\EE\left[\Lc(X_{t_{k+1}})\mid X_{t_k}\right] =\Lc(X_{t_k}) + \EE \left[   \Big(G(X_{t_k})\Big)^h \langle \nabla G(X_{t_k}) \cdot Z_{t_k} \rangle \right]+ o\left(  \Big(G(X_{t_k})\Big)^h \|  \nabla G(X_{t_k})\|\right)
\]
The assertion follows along the same lines as before.

\vspace{0.5 cm}

\profof{Corollary \ref{approx}}

 Under i.i.d. arrivals and static
constraints (i.e. when $\Hc=\Xc$) we can select  weak potential,  $G(X)$, as a Lyapunov 
function, then: 

From (\ref{2orderexp})
 and (\ref{nuova3}) we have: 
 \begin{multline*}
 \EE\left[G\Big(X_t +A_t-D_t^{\gmax}(I-R)\Big) \mid X_t \right]- G(X_t) \\ \stackrel{(\ref{appproxGdrift})}{=}
 \langle \nabla G(X_t) \cdot 
 \Lambda - D_t^{\gmax}(I-R)\rangle 
+o(\| \nabla G(X_t)\|) 
 \end{multline*}
with $ \langle \nabla G(X_t) \cdot \Lambda - D_t^{\gmax}(I-R)\rangle \stackrel{(\ref{drift})}{\le} - \epsilon \|
\nabla G(X_t)\| $ for sufficiently large $\|X_t\|$ and  an appropriate $\epsilon>0$.   

 Now again from (\ref{appproxGdrift}), substituting $D^{\imp}$ to $D^{\gmax}$ we have:
 \[
 \EE[G\Big(X_t +A_t-D^{\imp}(I-R)\Big) \big| X_t ]- G(X_t) 
 = \langle \nabla G(X_t) \cdot \Lambda - D^{\imp}_t(I-R) \rangle +o(\| \nabla G(X_t)\|) 
\]
 and by assumption: 
 \[
 \langle \nabla G(X_t) \cdot D^{\imp}_t(I-R)\rangle = \langle \nabla G(X_t) \cdot
 D^{\gmax}_t(I-R)\rangle 
 +o(\| \nabla G(X_t)\|)
 \]
Combining the two, we have:
 \begin{multline*}
 \EE\left[G\left(X_t +A_t-D^{\imp}(I-R)\right) \mid X_t \right]- G(X_t) =
 \langle \nabla G(X_t) \cdot \Lambda- D^{\imp}_t(I-R)\rangle + o(\| \nabla G(X_t)\|)  \\
 = \langle \nabla G(X_t) \cdot \Lambda - D^{\gmax}_t(I-R)\rangle +o(\| \nabla G(X_t)\|)
 \end{multline*}
 Thus
 \[
 \EE\left[G\left(X_t +A_t-D^{\imp}(I-R)\right) \big| X_t \right]- G(X_t) \\ 
 \le - \epsilon'\| \nabla G(X_t)\|
\]
for sufficiently large $\|X_t\|$  and $\epsilon'<\epsilon$.
 $\| \nabla G(X) \|$-stability for the system of queues follows.
 The proof in the case in which $G(X)$ is a strong potentials follows exactly along the same
 lines.
 Furthermore by adopting $\Lc(X)= \frac{1}{h+1} [G(X)]^{h+1}$ as a Lyapunov function
 and acting as before, the
 stability criterion can be strengthened.

\vspace{0.5 cm}

\profof{Theorem \ref{memory}}  

\proof 
We recall that in this case the space state of the
DTMC is $Y_t=[X_t, D^{\mem}_{t}]$. 
We select the follwing Lyapunov function: $$\Lc(Y_t)=\Lc(X_t,D^{\mem}_t)= \Lc_1(X_t)+ 
\Lc_2(X_t,D^{\mem}_t),$$ 
with 
$$\Lc_1(X_t)= G(X_t)$$ and 
\[
\Lc_2(X_t,D^{\mem}_t)
=\left(\langle \nabla G(X_t) \cdot
\Big(D_t^{\gmax}- D^{\mem}_t\Big)(I-R) \rangle \right)^\beta
\]
 where $\beta>1$ is given by  (\ref{asymptder-memory});
Observe that $\Lc_2(X_t,D^{\mem}_t)$ is well defined  because $\langle \nabla G(X_t) \cdot (D_t^{\gmax}- D^{\mem}_t)(I-R) \rangle \ge 0$
by construction.

Now we are  going to  show that  drift condition \eqref{outside-F} of Theorem \ref{th4} is satisfied.
By taking first order Taylor expansion  of $\Lc_1(X_{t+1})$ centered in $X_t$, we get:
\[
\Lc_1(X_{t+1}) =  G(X_{t+1})=   G(X_{t})+ \langle \nabla G(X_t) \cdot A_t- D^{\mem}
_t(I-R) \rangle + R^2_G(X, A_t- D^{\mem}_t(I-R))
\]
and recalling that  $R^2_G(X, Z)=  YH_h(X_t + \alpha Z) Z^T=  O(\|H_G(X_t)\|) $ for any  bounded vector $Z$,  in light of  (\ref{asymptderadd}),
we obtain:
$$\EE[\Lc_1(X_{t+1})-\Lc_1(X_{t})\mid Y_t] = \EE\left[\langle \nabla G(X_t) \cdot
 A_t- D^{\mem}_t(I-R) \rangle \mid Y_t \right] +  O(\|H_G(X_t)\|)$$


Now
\begin{multline*}
 \EE\Big[\langle \nabla G(X_t)\cdot A_t- D^{\mem}_t(I-R) \rangle \mid Y_t  \Big]=
 \langle \nabla G(X_t)\cdot  \Lambda-D^{\mem}_t(I-R) \rangle 
 \\= \langle \nabla G(X_t)\cdot \Lambda- (D_t^{\gmax} - D_t^{\gmax}- D^{\mem}_t)(I-R) \rangle 
 \\= \langle \nabla G(X_t)\cdot (D_t^{\gmax}- D^{\mem}_t)(I-R) \rangle + \langle \nabla G(X_t)\cdot
\Lambda- D_t^{\gmax}(I-R) \rangle  \\ 
 \stackrel{(\ref{drift})}{\le}  (\Lc_2(X_t,D^{\mem}_t))^{1/\beta} - \epsilon (\|\nabla G(X_t)\|) 
\end{multline*}
for an appropriate $\epsilon>0$. Indeed  by assumption $\Lambda(I-R)^{-1}$ lies in
the interior of $\Dc$.

Thus:
\begin{multline}
\EE[\Lc_1(X_{t+1},D^{\mem}_{t+1})-\Lc_1(X_{t},D^{\mem}_t)\mid Y_t] \\  \le
(\Lc_2(X_t,D^{\mem}_t))^{1/\beta} - \epsilon
(\|\nabla G(X_t)\|)+ O(\|H_G(X_t)\|).
\label{driftL1}
\end{multline}

Focusing instead on $\Lc_2(X_t,D^{\mem}_t)$,
we suppose for the moment $D_t\neq D^{\gmax}_t$: 
\begin{multline*}
\EE\Big[ \Lc_2(X_{t+1}, D^{\mem}_{t+1}) \Big| Y_t \mbox{ with } D^{\mem}_t\neq
D^{\gmax}_t\Big] \\ =
\EE\Big[\Lc_2(X_{t+1}, D^{\mem}_{t+1}) \Big| Y_t  \mbox{ with } D^{\mem}_t\neq D^{\gmax}_t,  D^{\mem}_{t+1}\neq
D^{\gmax}_{t+1} \Big] \cdot \\
 \Prob \big\{ D_{t+1}^{\mem}\neq D^{\gmax}_{t+1}\Big| Y_t \mbox{ with } D_t^{\mem}\neq D^{\gmax}_t \Big\}
\\ +
\EE\Big[\Lc_2(X_{t+1}, D_{t+1}^{\mem})\Big| Y_t  \mbox{ with } D^{\mem}_t\neq D^{\gmax}_t,  D^{\mem}_{t+1}=
D^{\gmax}_{t+1}\Big] 
 \cdot \\  \Prob\Big\{ D^{\mem}_{t+1}= D^{\gmax}_{t+1} \mid Y_t  \mbox{ with } D^{\mem}_t\neq D^{\gmax}_t \Big\} \\
\le \EE\Big[\Big(\langle \nabla G(X_{t+1}) \cdot (D_{t+1}^{\gmax}- D^{\mem}_{t+1})(I-R) \rangle \Big)^\beta \Big| Y_t  \mbox{ with }
\\ D^{\mem}_t\neq D^{\gmax}_t, D^{\mem}_{t+1} \neq D^{\gmax}_{t+1}\Big](1-\delta)
 \end{multline*}
 where the last inequality comes from the fact that by construction: 
 \begin{multline*}
 \EE\big[\Lc_2(X_{t+1},D_{t+1}^{\mem})\mid Y_t  \mbox{ with }  D_t^{\mem}\neq D^{\gmax}_{t}, 
 D^{\mem}_{t+1}= D^{\gmax}_{t+1}\big]
 \\=\EE\Big[\big(\langle \nabla G(X_{t+1}) \cdot (D_{t+1}^{\gmax}- D^{\gmax}_{t+1})(I-R) \rangle \big)^\beta\Big]=0
 \end{multline*}
 while $ \Prob\{D^{\mem}_{t+1}\neq D^{\gmax}_t\mid X_t, D_t^{\mem}\neq D^{\gmax}_t\}\le 1-\delta$.
 
 Now since our scheme guarantees that: 
 $\langle \nabla G(X_{t+1}) \cdot D^{\mem}_{t+1}(I-R) \rangle \;\; \ge \; \; \langle \nabla
 G(X_{t+1}) \cdot D_{t}^{\mem}(I-R)\rangle $ we can write:

\begin{multline*}
 \EE\Big[\Big(\langle \nabla G(X_{t+1}) \cdot (D_{t+1}^{\gmax}- D_{t+1}^{\mem})(I-R) \rangle \Big)^\beta
\Big| Y_t \mbox{ with }  D_t^{\mem}\neq D^{\gmax}_t, D^{\mem}_{t+1} \neq D^{\gmax}_{t+1}\Big](1-\delta)
\\  \le \EE\Big[\Big(\langle \nabla G(X_{t+1}) \cdot \big( D_{t+1}^{\gmax}- D^{\mem}_{t}\big) (I-R) \rangle \Big)^\beta \Big|
 Y_t  \mbox{ with } D^{\mem}_t\neq D_t^{\gmax} \Big] (1-\delta) \\ =
\EE\Big[\Big(\langle \nabla G(X_{t+1})  \cdot \big( D_{t+1}^{\gmax}-D_{t}^{\gmax}+D_{t}^{\gmax}-
 D_{t}^{\mem} \big)(I-R)\rangle \Big)^\beta \Big| 
 Y_t \mbox{ with }\\ D_t^{\mem}\neq D^{\gmax}_t\Big] (1-\delta)\\ =
\EE\Big[ \Big(\langle \nabla G(X_{t+1}) \cdot \big( D_{t+1}^{\gmax}-D_{t}^{\gmax}\big)(I-R)\rangle \\ 
 + \langle \nabla G(X_{t+1}) \cdot \big( D_{t}^{\gmax}-D^{\mem}_{t}\big)(I-R) \rangle \Big)^\beta \mid  Y_t \mbox{ with }\\ D_t^{\mem}\neq D^{\gmax}_t   \Big]  (1-\delta). 
\end{multline*} 
Expanding component-wise in Taylor series $\nabla G(X)$, we can write, similarly as before,
$\nabla G(X_{t+1})= \nabla G(X_t+A_t-D^{\mem}_t(I-R))=\nabla G(X_t)+O(\|H_G(X_t)\|)$ in light 
of (\ref{nuova2}) and of the fact that both
 $A_t$ and $D_t^{\mem}$ are bounded. Furthermore observe that by Lemma \ref{lemma2}:
 $\langle \nabla G(X_t) \cdot (D_{t+1}^{\gmax}-D_{t}^{\gmax})(I-R) \rangle \le o(\| \nabla G(X_t) \|) $; 
therefore we obtain:
\begin{multline*}
\EE[\Lc_2(X_{t+1},D^{\mem}_{t+1})\mid Y_t \mbox{ with } D^{\mem}_t\neq D^{\gmax}_t]  
\\ \le \Big(\langle \nabla
G(X_{t}) \cdot \big(D_{t}^{\gmax}- D^{\mem}_{t}\big)(I-R) \rangle (1-\delta)+o(\| \nabla G(X_t) \|)) \Big)^\beta
 \\ = \Lc_2(X_t,D^{\mem}_t)(1-\delta) +o\big(\Lc_2(X_t,D^{\mem}_{t})\big).
 \end{multline*} 
Thus:
\begin{multline}
\EE[\Lc_2(X_{t+1},D^{\mem}_{t+1})\mid Y_t  \mbox{ with } D^{\mem}_t\neq D^{\gmax}_t]- \Lc_2(X_{t},D^{\mem}_t) \\ \le
-\delta \Lc_2(X_t,D^{\mem}_{t}) +o\big(\Lc_2(X_t,D^{\mem}_{t})\big).
\label{driftL2}
\end{multline}
When $D^{\mem}_t= D^{\gmax}_t $, instead: 
\begin{multline}
 \EE\Big[\Lc_2(X_{t+1},D^{\mem}_{t+1})\mid Y_t \mbox{ with }  D^{\mem}_t= D^{\gmax}_t\Big]\\ 
 =  \EE\Big[\Big(\langle \nabla G(X_{t+1}) \cdot (D_{t+1}^{\gmax}- D_{t+1}^{\mem})(I-R) 
\rangle \Big)^\beta\Big]\\  \stackrel{\eqref{propincr}}{\le}
\EE\Big[\Big(\langle \nabla G(X_{t+1}) \cdot (D_{t+1}^{\gmax}- D_{t}^{\gmax})(I-R) 
\rangle \Big)^\beta\Big] \\
\le  O( \|H_G(X_t)\|^\beta )).
\label{driftL2a}
\end{multline}

Combining together (\ref{driftL1}) and (\ref{driftL2}) or (\ref{driftL2a}), 
(we recall that $\Lc_2(X_{t},D^{\mem}_{t})=0$ if$D^{\mem}_t= D^{\gmax}_t$) we obtain:
\begin{multline}
\EE[\Lc(X_{t+1},D^{\mem}_{t+1})\mid 
Y_t ] \\ \le - \epsilon (\|\nabla G(X_t)\|)+
 (\Lc_2(X_t,D^{\mem}_t))^{1/\beta} -\delta \Lc_2(X_t,D^{\mem}_t)+ o(\Lc_2(X_t,D^{\mem}_{t}))) 
 +  O(\|H_G(X_t) \|^\beta) \\ 
= - \epsilon (\|\nabla G(X_t)\|)  -\delta \Lc_2(X_t,D^{\mem}_t) )+
 o(\| \nabla G(X_t) \|  ) + o(\Lc_2(X_t,D^{\mem}_{t}))) 
\label{driftL}
\end{multline}
in light of (\ref{asymptder-memory})  (i.e., of the fact  that 
$\beta$ is selected in such a way to guarantee that $\|H_G(X_t) \|^\beta= o(\|\nabla G(X_t)\|)$).
Thus for a sufficiently large $b>0$, we can claim that:
\[
\EE[\Lc(X_{t+1},D^{\mem}_{t+1})\mid Y_t]- \Lc (X_{t},D^{\mem}_t)\le - \epsilon' (\|\nabla G(X_t)\|)
\]
for any $Y_t$, such that $\|X_t\|>b$ and for
 any $\epsilon'<\epsilon$.

$\|\nabla G(X)\|$-stability for the system of queues follows, since for any
$Y_t:\|X_t\|\le b$, $\Lc(Y_{t+1})- \Lc(Y_{t})$ is bounded, as immediate consequence of the fact that $\| X_{t+1}-X_t\|$ is bounded.

The stability criterion can be strengthened. 
For any $h\in \enne$, we can prove that the system of queues is $\|X^h\|$-stable 
under any admissible arrival vector, 
by selecting the Lyapunov function 
$\Lc'(Y_t)=\Lc'(X_t,D_t^{\mem})= \frac{1}{h+1}\big(\Lc(X_t,D_t^{\mem})\big)^{h+1}=\frac{1}{h+1}\big(\Lc_1(X_t)+ 
\Lc_2(X_t,D_t^{\mem})\big)^{h+1}=\frac{1}{h+1}\big(G(X_t)+\left(\langle \nabla G(X_t) \cdot
(D_t^{\gmax}- D^{\mem}_t)(I-R) \rangle \right)^\beta \big)^{h+1}$. The derivation 
proceeds along the same lines as the proof of Corollary \ref{strong-stab}, essentially showing that 
$E\big[\Lc'(Y_{t+1})\mid Y_t\big]-\Lc'(Y_t)\approx 
\big(\Lc(X_t,D_t^{\mem})\big)^{h}\big(E\big[\Lc(X_{t+1},D_{t+1}^{\mem})\mid Y_t\big]-\Lc(X_t,D_t^{\mem})\big) \le - \epsilon (\|\nabla G(X_t)\|)\big(\Lc(X_t,D_t^{\mem})\big)^{h}$, 
for any $Y_t: \|X_t\|>b$ with a sufficiently large $b>0$, and for a sufficiently small $\epsilon>0$. 

\vspace{0.5 cm}

\profof{Theorem \ref{memory-strong}} 

This proof combines arguments already applied in the proofs of 
 Theorem \ref{max-scalar-theo-sh-strong} and Theorem \ref{memory}.
First, we observe that the
state of the DTMC that describes system evolution is given by vector: $Y_t=[X_t,S_t, S^M_t]$, 
where $S_t=[S^A_t,S^D_t]\in \Sc^A \times \Sc^D$ represents the 
dynamics of exogenous arrivals, and service constraints, 
while $S^{M}_t$ provides additional information about
the memory state of the scheduling algorithm; such extra information correspond to 
 the set of departure vectors $D^M_t(S^D)$,  $\forall S^D \in \Sc^D$ ,  memorized by the scheduling algorithm.
We recall that $D^M_t(S^D)$ is the departure vector 
 employed at the last occurrence of state $S^D$ before $t$.

 We select a Lyapunov function with a similar structure as the one
 used in Theorem~\ref{memory}, however this time, things are made slightly
 more difficult by the fact that the memory of the scheme is significantly
 larger. Furthermore the stability properties of $Y_t$ are derived
 from those of the DTMC $Y_{t_k}$ obtained through the sub-sampling of $Y_t$, at instants
 $\{t_k \}$ in which the DTMC $S_{t_k}=S_0$, for a particular state $S_0$  (Corollary~\ref{th3a}).
Again we can claim that $\{t_k\}$ forms a sequence of non-defective
regeneration points for the system. 



In more detail,  the selected  Lyapunov function is:
$$\Lc(Y_t)=\Lc(X_t, S^M_t)= \Lc_1(X_t)+ \Lc_2(X_t,S^M_t), $$ with:
$$\Lc_1(X_t)= G(X_t)$$ and 
$$\Lc_2(X_t, S^M_t)= \sum_{S \in \Sc} \pi_S \left(\langle \nabla G(X_t) 
\cdot (D_t^{\gmax}(S^D)- D^M_t(S^D))(I-R) \rangle \right)^\beta,$$ 
where, once again, we recall that  $ D^M_t(S^D)$ represents the memorized departure vector at time $t$, which corresponds to $S^D$  (i.e.,  the service constraints component
 of state $S$), $\beta>1$  is specified by (\ref{asymptder-memory}) and $\pi_S$ is the steady state probability of the
 DTMC $S_t$ governing arrivals and dynamic constraint conditions.
 
In the remainder of the proof to simplify the notation we omit the dependency of 
the departure vector on current constraints conditions, writing $D^{\mem}_t$ 
instead of $D^{\mem}_t(S^D_t)$, and  $D^{\gmax}_t$ instead of $D^{\gmax}_t(S^D_t)$  whenever this can be done
without causing confusion.

Taking   the second order Taylor expansion of  $G(X_{t_{k+1}})$ centered in $X_{t_k}$ we get:
\begin{multline*}
\EE\left[\Lc_1(X_{t_{k+1}})-\Lc_1(X_{t_k})\mid Y_t\right] \\= \EE\left[ \sum_{t=t_k}^{t_{k+1}-1}
\langle \nabla G(X_{t_k}) \cdot A_t- D^{\mem}_t(I-R) \rangle \mid Y_{t_k} \right] +O(\|H_G(X_t)\|)
\end{multline*}
with:
\begin{multline*}
 \EE\Big[ \langle \nabla G(X_{t_k}) \cdot \sum_{t=t_k}^{t_{k+1}-1} A_t- D^{\mem}_t(I-R) \rangle \mid Y_{t_k} \Big]\\
 = \EE\Big[ \langle \nabla G(X_{t_k})\cdot  \sum_{t=t_k}^{t_{k+1}-1} A_t -
 (D_t^{\gmax}- D_t^{\gmax}+ D^{\mem}_t)(I-R)\rangle \mid Y_{t_k} \Big]\\
 = \EE\Big[ \langle \nabla G(X_{t_k})\cdot \sum_{t=t_k}^{t_{k+1}-1} 
 (D_t^{\gmax}-D^{\mem}_t)(I-R) \rangle \mid Y_{t_k} \Big]\\
 +\EE\Big[ \langle \nabla G(X_{t_k})\cdot \sum_{t=t_k}^{t_{k+1}-1} A_t -
 D_t^{\gmax}(I-R) \rangle \mid Y_{t_k} \Big] 
\end{multline*}
Now,
$\EE\Big[\sum_{t=t_k}^{t_{k+1}-1} \langle \nabla G(X_t)\cdot A_t -
 D_t^{\gmax}(I-R) \rangle \mid Y_{t_k} \Big] \le - \epsilon \EE[z_k]
 (\|\nabla G(X_t)\|)$,  from  (\ref{driftstrong-1})   and (\ref{driftstrong}) in  the proof of
Theorem~\ref{max-scalar-theo-sh-strong}.

While: 
\begin{multline*}
 \EE\Big[ \langle \nabla G(X_{t_k})\cdot \sum_{t=t_k}^{t_{k+1}-1} 
 (D_t^{\gmax}-D^{\mem}_t)(I-R) \rangle \mid Y_{t_k} \Big]\\
 \EE\Big[ \langle \nabla G(X_{t_k})\cdot \sum_{t=t_k}^{t_{k+1}-1} 
 (D_t^{\gmax}-D^{\mem}_t + D_{t_k}^{\gmax}(S^D_t) -D^{M}_{t_k}(S^D_t)
-D_{t_k}^{\gmax}(S^D_t) +D^{M}_{t_k}(S^D_t) 
)(I-R) \rangle \mid Y_{t_k} \Big]
 \\ \le  \EE\Big[ \langle \nabla G(X_{t_k})
 \cdot \sum_{t=t_k}^{t_{k+1}-1} 
 (D_{t_k}^{\gmax}(S^D_t)-D^{M}_{t_k}(S^D_t))(I-R)\rangle \mid Y_{t_k}\Big]+O(\|H_G(X_t)\|)
\end{multline*}
where we recall that: $$D_{t_k}^{\gmax}(S^D_t)=\argmax_{\DSXA}\langle \nabla G(X_{t_k})\cdot D(I-R)\rangle $$ and $D^{M}_{t_k}(S^D_t)$ is the departure vector memorized by the scheduling 
at time $t_k$, which corresponds to state $S^D_t$.
Observe, indeed,  that   
$\EE\Big[ \langle \nabla G(X_{t_k})\cdot \sum_{t=t_k}^{t_{k+1}-1}  (D_t^{\gmax}-D^{\gmax}_{t_k}(S^D_t))(I-R) \rangle \mid Y_{t_k} \Big]\le o(\| \nabla G(X_{t_k})\|) $   from  Lemma \ref{lemma2} and 
$\EE\Big[ \langle \nabla G(X_{t_k})\cdot \sum_{t=t_k}^{t_{k+1}-1} D^{M}_{t_k}(S^D_t)  -D_t^{\mem})(I-R) \rangle \mid Y_{t_k} \Big]= 
\EE\Big[ \sum_{t=t_k}^{t_{k+1}-1} \langle \nabla G(X_{t})\cdot  D^{M}_{t_k}(S^D_t)  -D_t^{\mem})(I-R) \rangle \mid Y_{t_k} \Big] +O(\|H_G(X_t)\|)$, 
with  $\langle \nabla G(X_{t})\cdot (D^{M}_{t_k}(S^D_t)  -D_t^{\mem} )(I-R) \rangle  \le O(\|H_G(X_t)\|) $ 
as consequence of \eqref{propincrstrong} ,  (\ref{nuova2a})  and the fact that polynomial moments of $t_ {k+1}-t_k$ are finite (and thus also moments of  $t-t_k$ 
with $t \in \{t_k, \cdots, t_{k+1}-1 \}$).

Now:
\begin{multline*}
\EE\Big[ \langle \nabla G(X_{t_k})\cdot \sum_{t=t_k}^{t_{k+1}-1} 
 (D_{t_k}^{\gmax}(S^D_t)-D^{M}_{t_k}(S^D_t))(I-R)\rangle \mid Y_{t_k} \Big]
\\=
\EE[z_k]\sum_{S\in \Sc} \pi_S \langle \nabla G(X_{t_k})\cdot 
 (D_{t_k}^{\gmax}(S^D)-D^M_{t_k}(S^D))(I-R) \rangle \\
=
\EE[z_k]\Lc_2(X_{t_k},S^M_{t_k}))^{1/\beta}
\end{multline*}

Thus:
\begin{equation}
\EE[\Lc_1(X_{t_{k+1}})-\Lc_1(X_{t_k})\mid Y_{t_k}] \le\
\EE[z_k](\Lc_2(X_{t_k},S^M_{t_k}))^{1/\beta}- 
\epsilon \|(\nabla G(X_{t_k}))\| + O(\|H_G(X_{t_k})\|)
\label{driftL1-strong}
\end{equation}


Focusing instead on $\Lc_2(X_{t_k},S^M_{t_k})$,
first observe that   $\forall t \in \{t_k,\cdots, t_{k+1}-1\} $,  for any $S^D\in \Sc^D$, 
we have: 

\begin{multline}
\EE\Big[\Big(\langle \nabla G(X_{t_{k+1}}) \cdot (D_{t_{k+1}}^{\gmax}(S^D)- D_{t_{k+1}}^{M}(S^D))(I-R) 
\rangle \Big)^\beta \mid Y_t \mbox{ with } D_{t}^{M}(S^D)= D_{t}^{\gmax}(S^D) \Big] \\ =
\EE\Big[\Big(\langle \nabla G(X_{t_{k+1}}) \cdot (D_{t_{k+1}}^{\gmax}(S^D)- D_{t}^{\gmax}(S^D))(I-R) 
\rangle \\  + \langle \nabla G(X_{t_{k+1}}) \cdot (D_{t}^{\gmax}(S^D)- D_{t_{k+1}}^{M}(S^D))(I-R) 
\rangle \Big)^\beta  \mid Y_t \mbox{ with } D_{t}^{M}(S^D)= D_{t}^{\gmax}(S^D) \Big] \\ = 
\EE \Big[\Big(\langle \nabla G(X_{t_{k+1}}) \cdot (D_{t_{k+1}}^{\gmax}(S^D)- D_{t}^{\gmax}(S^D))(I-R) 
\rangle + \langle \nabla G(X_{t_{k+1}}) \cdot (D_{t}^{M}(S^D)- D_{t_{k+1}}^{M}(S^D))(I-R) 
\rangle \big)^\beta \Big]
 \\ \le \EE\Big[\Big(\langle \nabla G(X_{t_{k+1}}) \cdot
 (D_{t_{k+1}}^{\gmax}(S^D)- D_{t}^{\gmax}(S^D))(I-R) 
\rangle + O\big( \|H_G (X_{t_{k+1}})\|\big) \Big)^\beta \Big] 
\label{omicromlstrong1}
\end{multline}
because by construction $\langle \nabla G(X_{t_{k+1}}) \cdot (D_{t}^{M}(S^D)- D_{t_{k+1}}^{M}(S^D))(I-R) \le O\big( \|H_G (X_{t_{k+1}})\|\big) $ 
as consequence of \eqref{propincrstrong} ,  (\ref{nuova2a})  and the fact that polynomial moments of $t_ {k+1}-t_k$ are finite (and thus also moments of  $t_ {k+1}-t$ 
with $t \in \{t_k, \cdots, t_{k+1}-1 \}$);    Moreover observe that $(x)^\beta$ is monotone increasing (as its argument is surely positive).  Now:
\begin{multline}
 \EE\Big[\Big(\langle \nabla G(X_{t_{k+1}}) \cdot (D_{t_{k+1}}^{\gmax}(S^D)- D_{t}^{\gmax}(S^D))(I-R) 
\rangle \Big) \Big] \\  =
\EE \Big[\Big( \langle \nabla G(X_{t}) \cdot (D_{t_{k+1}}^{\gmax}(S^D)- D_{t}^{\gmax}(S^D))(I-R) 
\rangle  \\ +
 O\big( \| H_G(X_{t})\| \big)\Big)\Big]
 \le  O\Big( \|H_G (X_{t}) \|\Big)=  O\Big( \|H_G (X_{t_{k+1}}) \|\Big)
\label{omicromlstrong2}
\end{multline}
where   first equality is again a direct consequence of  (\ref{nuova2a}) and the fact that polynomial moments of $t_ {k+1}-t_k$ are finite;
the following inequality is a consequence of the fact that 
$\langle \nabla G(X_{t}) \cdot (D_{t_{k+1}}^{\gmax}(S^D)- D_{t}^{\gmax}(S^D))(I-R)  \rangle =O\big( \| H_G(X_{t})\| \big)$.
Thus combining (\ref{omicromlstrong1}) and (\ref{omicromlstrong2})  $\forall t \in \{t_k. \cdots, t_{k+1}-1\}$, 
we get: 
\begin{multline}
\EE\Big[\Big(\langle \nabla G(X_{t_{k+1}}) \cdot (D_{t_{k+1}}^{\gmax}(S^D)- D_{t_{k+1}}^{M}(S^D))(I-R) 
\rangle \Big)^\beta\Big] \mid  D_{t}^{M}(S^D)= D_{t}^{\gmax}(S^D) \Big] =O\Big( \|H_G (X_{t}\|\Big)^\beta
\label{omicromlstrong3}
\end{multline}

Now:
\begin{multline}
\EE \Big[\Big(\langle \nabla G(X_{t_{k+1}}) \cdot (D_{t_{k+1}}^{\gmax}(S^D)- D_{t_{k+1}}^{M}(S^D))(I-R) 
\rangle \Big)^\beta \mid Y_{t_k}  \mbox{ with }  D_{t_{k}}^{M}(S^D)\neq D_{t_{k}}^{\gmax}(S^D) \Big] = \\  
\EE\Big[\Big(\langle \nabla G(X_{t_{k+1}}) \cdot (D_{t_{k+1}}^{\gmax}(S^D)- D_{t_{k+1}}^{M}(S^D))(I-R) 
\rangle \Big)^\beta \mid Y_{t_k},  D_{t}^{M}(S^D)\neq D^{\gmax}_t (S^D) \; \forall t \in \{t_k, \cdots,  t_{k+1}\}  \Big] \cdot \\ 
\Prob  \left\{  D_{t}^{M}(S^D)\neq D^{\gmax}_t (S^D) \; \forall t \in \{t_k+1,\cdots, t_{k+1}\} \mid Y_{t_k} \mbox{ with } D_{t_k}^{M}(S^D)\neq D^{\gmax}_{t_k} (S^D)   \right\} +\\
\EE\Big[\Big(\langle \nabla G(X_{t_{k+1}}) \cdot (D_{t_{k+1}}^{\gmax}(S^D)- D_{t_{k+1}}^{M}(S^D))(I-R) 
\rangle \Big)^\beta \mid Y_{t_k} \mbox{ with } D_{t_k}^{M}(S^D)\neq D^{\gmax}_{t_k} (S^D),\\   \exists t \in \{t_k+1,\cdots, t_{k+1}\} : \;  D_{t}^{M}(S^D) =  D^{\gmax}_t (S^D)    \Big] \cdot \\ 
\Prob \left\{   \exists t \in \{t_k+1,\cdots, t_{k+1}\} : \;  D_{t}^{M}(S^D) =  D^{\gmax}_t (S^D) \mid  Y_{t_k}  \mbox{ with } D_{t_k}^{M}(S^D)\neq D^{\gmax}_{t_k} (S^D)   \right\} =\\
\EE\Big[\Big(\langle \nabla G(X_{t_{k+1}}) \cdot (D_{t_{k+1}}^{\gmax}(S^D)- D_{t_{k+1}}^{M}(S^D))(I-R) 
\rangle \Big)^\beta \mid Y_{t_k},  D_{t}^{M}(S^D)\neq D^{\gmax}_t (S^D) \; \forall t \in \{ t_k,\cdots, t_{k+1} \}  \Big]  
 (1- \delta \hat{\pi}_S) + \\
O( \| H_G(X_{t÷k})\| )
\label{driftL2bstrong}
\end{multline}
where $\hat{\pi}_{S^D}$ denotes the probability that one of  states $S$,  whose service constraint component is equal to  $S^D$, is visited in $\{ t_k+1,\cdots, t_{k+1}\}$ (i.e.,
 between to following visits to state $S_0$).
Indeed  observe that by construction $\Prob  \left\{  D_{t}^{M}(S^D)\neq D^{\gmax}_t (S^D) \; \forall t \in \{t_k+1,\cdots, t_{k+1}\} \mid  Y_{t_k}  \mbox{ with } D_{t_k}^{M}(S^D)\neq D^{\gmax}_{t_k} (S^D)  
 \right\} \le 1- \delta \hat{\pi}_{{S}^D}$ since  $D_{t}^{M}(S^D) = D_{t}^{\gmax}$  with a probability  greater than $\delta$,
provided that one of the state $S$  with  service constraints  equal to    $S^D$   has been visited at least once in $\{t_k+1,\cdots, t_{k+1}\}$;  Moreover we can apply  (\ref{omicromlstrong3}) to 
$\EE\Big[\Big(\langle \nabla G(X_{t_{k+1}}) \cdot (D_{t_{k+1}}^{\gmax}(S^D)- D_{t_{k+1}}^{M}(S^D))(I-R) 
\rangle \Big)^\beta \mid   Y_{t_k}  \mbox{ with }  D_{t_k}^{M}(S^D) \neq  D^{\gmax}_{t_k} (S^D),  \exists t \in \{t_k+1,\cdots, t_{k+1}\} : \;  D_{t}^{M}(S^D) =  D^{\gmax}_t (S^D)   \Big]$.

At last 
\begin{multline}
\EE\Big[\Big(\langle \nabla G(X_{t_{k+1}}) \cdot (D_{t_{k+1}}^{\gmax}(S^D)- D_{t_{k+1}}^{M}(S^D))(I-R) 
\rangle \Big)^\beta \\ \mid Y_{t_k},  D_{t}^{M}(S^D)\neq D^{\gmax}_t (S^D)\;\forall t \in  \{t_k,\cdots, t_{k+1}\}  \Big]   \\  \le
\EE\Big[\Big(\langle \nabla G(X_{t_{k+1}}) \cdot (D_{t_{k+1}}^{\gmax}(S^D)- D_{t_{k}}^{M}(S^D))(I-R) 
\rangle  + O\big( \|H_G (X_{t})\|\big) \Big)^\beta ) \Big]  \\  \le
\EE\Big[\Big(\langle \nabla G(X_{t_{k}}) \cdot (D_{t_{k}}^{\gmax}(S^D)- D_{t_{k}}^{M}(S^D))(I-R) 
\rangle  + O\big( \|H_G (X_{t})\| \big) \Big)^\beta \Big] 
\label{driftL2bstrong2}
\end{multline}
where the  first inequality  derive from (\ref{propincrstrong}) (\ref{nuova2a})  and the fact that polynomial moments of $t_ {k+1}-t_k$ are finite; while the following inequality  can be obtained expanding  in Taylor series  $G(X_{t_{k+1}})$ around $X_{t_k}$  and observing that by construction 
$\nabla G(X_{t_{k}}) \cdot (D_{t_{k}}^{\gmax}(S^D)- D_{t_{k+1}}^{\gmax}(S^D))(I-R) \le  O\big( \|H_G (X_{t})\|\big) $. 

Thus  combining (\ref{driftL2bstrong}) and  (\ref{driftL2bstrong2})  we get:
\begin{multline}
\EE \Big[\Big(\langle \nabla G(X_{t_{k+1}}) \cdot (D_{t_{k+1}}^{\gmax}(S^D)- D_{t_{k+1}}^{M}(S^D))(I-R) 
\rangle \Big)^\beta \mid  D_{t_{k}}^{M}(S^D)\neq D_{t_{k}}^{\gmax}(S^D) \Big]   \\ \le
\EE\Big[\Big(\langle \nabla G(X_{t_{k}}) \cdot (D_{t_{k}}^{\gmax}(S^D)- D_{t_{k}}^{M}(S^D))(I-R) (1- \delta \hat{\pi}_{S^D})
\rangle   \Big)^\beta \Big] + o\left(\Big(\langle \nabla G(X_{t_{k}}) \cdot \big(D_{t_{k}}^{\gmax}(S^D)- D_{t_{k}}^{M}(S^D)\big)(I-R)  \rangle\Big)^\beta\right)
\label{driftL2bstrong-fin}
\end{multline}
Now  multiplying  for $\pi_S$  and summing over all the states and recalling (\ref{driftL2bstrong-fin}), (\ref{omicromlstrong3}),
and   $\nabla G(X_{t_{k}}) \cdot \big(D_{t_{k}}^{\gmax}(S^D)- D_{t_{k}}^{M}(S^D)\big)(I-R)   
\rangle=0$  when $D_{t_{k}}(S^D) =  D_{t_{k}}^{\gmax}(S^D)$,    we get:

\begin{multline}
\EE[\Lc_2(X_{t_{k+1}}, S^M_{t_{k+1}})\mid X_{t_k}, S^M_{t_k}]  \\  =
 \sum_S \pi_S  \EE \Big[\Big(\langle \nabla G(X_{t_{k+1}}) \cdot (D_{t_{k+1}}^{\gmax}(S^D)- D_{t_{k+1}}^{M}(S^D))(I-R) 
\rangle \Big)^\beta \mid  X_{t_k}, S^M_{t_k}  \Big]  \\ \le 
 \sum_S \pi_S(1- \delta \hat{\pi}_{S^D})  \Big[\Big(\langle \nabla G(X_{t_{k}}) \cdot (D_{t_{k}}^{\gmax}(S^D)- D_{t_{k}}^{M}(S^D))(I-R)   
\rangle   \Big)^\beta \Big] \\  +
o\Big(\sum_S\pi_S  \big(  \langle \nabla G(X_{t_{k}}) \cdot (D_{t_{k}}^{\gmax}(S^D)- D_{t_{k}}^{M}(S^D))(I-R)   
\rangle   \big)^\beta \Big)  + o(\|H_G(X_{t_{k}})\|^\beta)\\  \le
\EE[\Lc_2(X_{t_{k}}, S^M_{t_{k}})](1- \delta \min_S \pi_S\hat{ \pi}_{S^D})) + o(\Lc_2(X_{t_{k}}, S^M_{t_{k}})) +  o(\|H_G(X_{t_{k}})\|^\beta)
\label{driftL2bstrong-fin-fin}
\end{multline}
where  observe that $\min_S \pi_S\hat{ \pi}_{S^D}> 0$ since $S_t$ is a finite state ergodic  Markov Chain. 
At last, by combining (\ref{driftL2bstrong-fin-fin}) and (\ref{driftL1-strong}), 
 we can easily show that  drift condition (\ref{outside-sampledF}) is satisfied.


$\|\nabla G(X)\|$-stability for the system of queues easily follows, since 
 (\ref{inside-sampledF}) can be easily derived (as for the previous Theorems)
 from the following three facts: 
 i) $\Lc(Y)$ and is indefinitely continuously derivable, and thus bounded
 (along with its derivatives) over compact
 domains; ii) at any instant $t$ both arrival and departure vectors are
 bounded; iii) $\EE[(z_k)^h]$ are bounded for any $h>0$.

We further notice that the stability criterion can be strengthened.
For any $h\in \enne$, we can prove that the system of queues is $\|X^h\|$-stable 
under any admissible arrival vector, 
by selecting the Lyapunov function 
$\Lc'(Y_t)=\Lc'(X_t,D_t^{\mem})= \frac{1}{h+1}\big(\Lc(X_t,D_t^{\mem})\big)^{h+1}=\frac{1}{h+1}\big(\Lc_1(X_t)+ 
\Lc_2(X_t,D_t^{\mem})\big)^{h+1}$.


\end{sloppypar}

\begin{thebibliography}{99}

\bibitem{nostro-tit} M. Ajmone Marsan, E. Leonardi, M. Mellia, F. Neri, ``On the stability
of isolated and interconnected input-queueing switches under multi-class
traffic,'' {\em IEEE Trans. Inform. Theory}, vol. 51, no. 3, pp. 1167-1174,
March 2005.

\bibitem{sarkar} P. Chaporkar, K. Kar, S. Sarkar, ``Throughput Guarantees in
Maximal Scheduling in Wireless Networks,'' {\em Allerton Conference on Communication, Control and Computing},
2005.
 
\bibitem{dai-lin} J. G. Dai, W. Lin, ``Maximum Pressure Policies in Stochastic Processing
Networks'', Operations Research, vol. 53, 2005, pp. 197-218

\bibitem{walrand} A. Dimakis, J. Walrand, ``Sufficient conditions for stability of
longest-queue-first scheduling: Second-order properties using fluid
limits,'' {\em Adv. Appl. Probab.}, vol. 38, no. 2, pp. 505-521, 2006.

\bibitem{Erylmaz} A. Eryilmaz, R. Srikant, J. R. Perkins,
``Stable Scheduling Policies for Fading Wireless Channels'', {\em IEEE/ACM
Transactions on Netqorking} Vol.13,  n 2, pp. 411-424, 2005.

\bibitem{giaccone-shah} P. Giaccone, B. Prabhakar, D. Shah, ``Randomized-scheduling algorithms for high-aggregate bandwidth switches'', {\em IEEE J. Sel. Areas Commun.}, Vol 21, n. 4 pp. 546- 559, May 2003.

\bibitem{gupta-sig} 
G. Gupta, S. Sanghavi, N. Shroff, ``Node Weighted Scheduling'', 
{\em ACM SIGMETRICS, 2009}, 
 June 2009

\bibitem{keslassy} I. Kesslassy, N. Mckeown, ``Analysis of scheduling
algorithms that provide 100\% throughput in input-queued switches'', {\em Allerton Conference on Communication 
 Control and Computer}, October 2001.

\bibitem{kushner} H.J.Kushner, {\em Stochastic Stability and Control,}
Academic Press, 1967.

\bibitem{nostro-jacm} E. Leonardi, M. Mellia, F. Neri, M. Ajmone Marsan,
``Bounds on delays and queue lengths in input-queued
cell switches'', {\em Journal of ACM}, Vol. 50(4), pp. 520-550, 2003.

\bibitem{LPF} A. Mekkittikul, N. McKeown, ``A Practical Scheduling Algorithm to
Achieve 100\% Throughput in Input-Queued Switches'', { \em INFOCOM}, 
April 1998.

\bibitem{meyn}  S. Meyn. (2009), ``Stability and asymptotic optimality of generalized maxweight policies,''
{\em SIAM Journal on control and optimization}, no. 47, pp. 3259-3294.


\bibitem{modiano-shah} E. Modiano, D. Shah, G. Zussman, ``Maximizing
Throughput in Wireless networks via Gossiping'', {\em ACM SIGMETRICS 2006}.

\bibitem{neely-modiano} M. J. Neely, E. Modiano, C. E. Rohrs, ``Dynamic power allocation
and routing for time varying wireless networks,'', {\em IEEE J. Sel. Areas
Commun.}, vol. 23, no. 1, pp. 89-103, Jan. 2005.

 \bibitem{bambos} K. Ross, N. Bambos. ``Projective cone scheduling (PCS) algorithms for packet switches of maximal
throughput,'' {\em Transactions on Networking}, Vol. 17, n. 3, pp. 
976-989, June 2009. 
 
\bibitem{deva-damon1} D. Shah, D. J. Wischik, 
``Optimal scheduling algorithms for input-queued switches'', 
{\em INFOCOM}, April 2006.
 
\bibitem{deva-damon2} D. Shah, D. J. Wischik,
 ``The teleology of scheduling algorithms for switched networks under light
 load, critical load, and overload'', Technical report, available on line at: 
 {\url http://www.cs.ucl.ac.uk/staff/ ucacdjw/Research/netsched.html}.

\bibitem{shah-sig} D. Shah, J. Tsitsiklis, Y. Zhong, ``Qualitative Properties
of $\alpha$-Weighted Scheduling Policies'', {\em ACM SIGMETRICS 2010}.



\bibitem{tassiulas-ephremides} L. Tassiulas, A. Ephremides, ``Stability properties of constrained
queuing systems and scheduling policies for maximum throughput in
multi-hop radio networks'', {\em IEEE Trans. Automat. Contr.}, vol. 37, pp. 1936-1948, Dec. 1992.

 \bibitem{linear-complexity-tassiulas} L. Tassiulas, ``Linear complexity algorithms for maximum throughput
in radio networks and input queued switches'', {\em INFOCOM}, April 1998, 

  \bibitem{MMBP-Tassiulas} L. Tassiulas, ''Scheduling and performance limits of networks with constantly changing topology,'' {\em  IEEE Transactions on Information Theory}, vol.43, no.3, pp.1067,1073, May 1997.

\bibitem{srikant}
X. Wu, R. Srikant, ``Scheduling efficiency of distributed greedy
scheduling algorithms in wireless networks'', {\em INFOCOM},
April 2006.

\bibitem{tweedie-meyn}  S.P. Meyn and R.L. Tweedie (1993), {\em  Markov chains and stochastic stability}. Cambridge University Press, Second Edition 2009. 






\end{thebibliography}
\end{document}